%% file: main.tex
\documentclass[sigconf]{acmart}
\usepackage{xcolor}
\usepackage{amsmath}
\usepackage{bm}
\input{meta/copyright}
\input{meta/acronyms}
\input{meta/packages}
\usepackage{acmart-taps}

\copyrightyear{2026}
\acmYear{2026}
\setcopyright{cc}
\setcctype{by-nc-nd}
\acmConference[CHI '26]{Proceedings of the 2026 CHI Conference on Human Factors in Computing Systems}{April 13--17, 2026}{Barcelona, Spain}
\acmBooktitle{Proceedings of the 2026 CHI Conference on Human Factors in Computing Systems (CHI '26), April 13--17, 2026, Barcelona, Spain}
\acmPrice{}
\acmDOI{10.1145/3772318.3791254}
\acmISBN{979-8-4007-2278-3/2026/04}

\newcommand{\sys}{\texttt{GitAcademy}}

\newcommand{\inlinequote}[1]{``\textit{#1}''}

\newcommand{\frev}[1]{\textcolor{black}{#1}}

\newcommand{\rev}[1]{#1}

\newcommand{\sd}[1]{$SD=#1$}
\newcommand{\mean}[1]{$M=#1$}

\newcommand{\md}[1]{$MD=#1$}

\newcommand{\ORG}{organizing}
\newcommand{\STR}{strategizing}
\newcommand{\LE}{learning exchange}
\newcommand{\TP}{teaching peer}
\newcommand{\FP}{following peer}
\newcommand{\gitpushsolved}{exercise performance}

\begin{document}

\input{meta/authors}

\input{sections/00_abstract}

\input{meta/metadata}

\renewcommand{\shortauthors}{Bucher and Goswami et al.}

\begin{teaserfigure}
  \centering    
  \includegraphics[width=\textwidth]{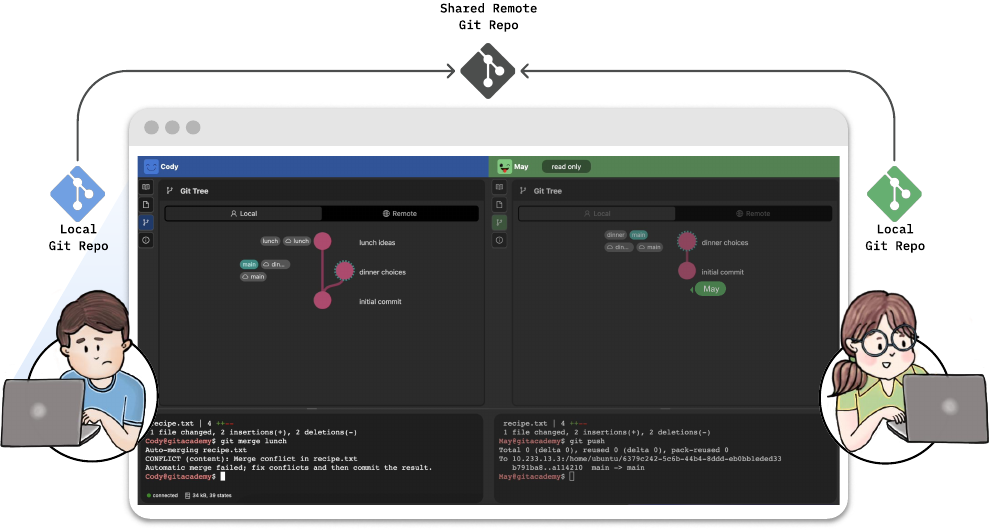}
  \Description{Diagram illustrating a Git collaboration platform between two users, Cody and May, with split-view awareness of each other’s workspace. The interface shows each user working in a local Git repository connected through a shared remote repository.}
  \caption{GitAcademy: An online learning platform that supports collaborative Git learning through split-view awareness, allowing learners to work together while maintaining individual context.}
  \label{fig:teaser}
\end{teaserfigure}
\maketitle

\input{sections/01_introduction}
\input{sections/02_related_work}
\input{sections/03_system_design}
\input{sections/05_evaluation}

\input{sections/06_results}

\input{sections/07_discussion}
\input{sections/08_conclusion}

\bibliographystyle{ACM-Reference-Format}
\bibliography{ref}
\input{sections/10_appendix}
\end{document}

%% file: meta/copyright.tex
\AtBeginDocument{%
  }

\setcopyright{acmlicensed}
\copyrightyear{2018}
\acmYear{2018}
\acmDOI{XXXXXXX.XXXXXXX}

\acmConference[Conference acronym 'XX]{Make sure to enter the correct
  conference title from your rights confirmation emai}{June 03--05,
  2018}{Woodstock, NY}
\acmISBN{978-1-4503-XXXX-X/18/06}



%% file: meta/acronyms.tex
\usepackage[nolist]{acronym}
\begin{acronym}
    \acro{HCI}{Human-Computer Interaction}
    \acro{CSCW}{Computer-Supported Collaborative Work}
    \acro{CS}{Computer Science}
\end{acronym}

%% file: meta/packages.tex
\usepackage{multirow}
\usepackage{graphicx}
\usepackage{amsmath}
\usepackage{bm}
\usepackage{microtype}
\usepackage{subcaption}
\usepackage{dirtytalk}
\usepackage{balance} 
\usepackage{enumitem}
\usepackage{color}
\usepackage{url}

%% file: meta/authors.tex

\author{Joel Bucher}
\authornote{These authors contributed equally to this work.}
\orcid{0009-0008-2671-1835}
\email{bucherjo@student.ethz.ch}
\affiliation{%
  \institution{ETH Zurich}
  \city{Zurich}
  \country{Switzerland}}

\author{Lahari Goswami}
\authornotemark[1]
\orcid{0000-0002-8975-5885}
\email{lahari.goswami@inf.ethz.ch}
\affiliation{%
  \institution{ETH Zurich}
  \city{Zurich}
  \country{Switzerland}}

\author{Sverrir Thorgeirsson}
\orcid{0000-0002-4455-7551}
\email{sverrir.thorgeirsson@inf.ethz.ch}
\affiliation{%
  \institution{ETH Zurich}
  \city{Zurich}
  \country{Switzerland}}

\author{April Yi Wang}
\orcid{0000-0001-8724-4662}
\email{april.wang@inf.ethz.ch}
\affiliation{%
  \institution{ETH Zurich}
  \city{Zurich}
  \country{Switzerland}}


%% file: sections/00_abstract.tex
\begin{abstract}
Git is widely used for collaborative software development, but it can be challenging for newcomers. 
While most learning tools focus on individual workflows, Git is inherently collaborative. 
We present GitAcademy, a browser-based learning platform that embeds a full Git environment with a split-view collaborative mode: learners work on their own local repositories connected to a shared remote repository, while simultaneously seeing their partner's actions mirrored in real time.
This design is not intended for everyday software development, but rather as a training simulator to build awareness of distributed states, coordination, and collaborative troubleshooting.
In a within-subjects study with 13 pairs of learners, we found that the split-view interface enhanced social presence, supported peer teaching, and was consistently preferred over a single-view baseline, even though performance gains were mixed. 
We further discuss how split-view awareness can serve as a training-only scaffold for collaborative learning of Git and other distributed technical systems.
\end{abstract}

%% file: meta/metadata.tex

\title{Git Takes Two: Split-View Awareness for Collaborative Learning of Distributed Workflows in Git}

\ccsdesc[500]{Human-centered computing~Collaborative interaction}
\ccsdesc[500]{Human-centered computing~Collaborative and social computing systems and tools}
\ccsdesc[500]{Human-centered computing~Empirical studies in HCI}
\keywords{Git, collaborative learning, educational groupware technology}


%% file: sections/01_introduction.tex
\section{Introduction} \label{intro}

In software engineering teams, Git is the most dominant version control system used to plan, manage, and coordinate changes to codebases across members \cite{stackoverflow2022}. 
However, despite its widespread adoption, learning Git remains a challenge for beginners and even experienced developers alike~\cite{rosso_jackson_2013, Church2014, Isomottonen2014}. 
For beginners, its extensive command set and high interactivity between components reduce Git's overall learnability~\cite{Hynes2015}. 
For instance, new learners often struggle with applying developer workflows, resolving merge conflicts, and distinguishing between local and remote repositories~\cite{eraslan_2020}. 
Even experienced developers face difficulties in using the full spectrum of Git commands and its underlying information model~\cite{yang2022}, and rely on familiar workflows with a small set of commands due to uncertainty and risk aversion~\cite{Church2014}.

Students are typically introduced to version control with Git through standalone computer science (CS) courses \cite{missingsemester2020} or in group projects in CS~\cite{tushev2020using}. 
Many new learners also rely on documentation and tutorials for self-study. 
In professional settings, new developers receive training on standard workflows in associated projects, mentoring from team leads and experienced peers, and knowledge sharing in communities~\cite{Church2014}. 
Although these pedagogical trainings help learners become familiar with Git, in practice most learners often use and acquire Git commands and concepts on demand~\cite{yang2022}, in real-life project contexts. 
This form of learning can be high stake: executing incorrect commands or using them in wrong contexts can lead to serious unintended consequences and data loss~\cite{Hynes2015}.
Crucially, learning Git is not only about mastering commands but also about coordinating with others. 
GUI tools can simplify command syntax, but they do not eliminate the conceptual difficulty of reasoning about distributed state across collaborators. 
Push, pull, and commit may appear straightforward in isolation, yet in collaborative contexts they interact with diverging histories, remotes, and timing. 
Novices therefore stumble less on syntax than on state reasoning: What exactly am I pushing? What did my partner fetch? Whose history rewrites whom?
Our premise is that effective Git learning must train this coordination layer -- not just command-level fluency.

However, opportunities to practice such coordination remain limited. 
Git was designed to facilitate asynchronous and distributed collaboration, yet most students encounter it only in tightly scoped classroom exercises or ad hoc project work. 
Research has shown that realistic project-based courses help students recognize the value of version control systems and better understand Git's collaborative use~\cite{Isomottonen2014}. 
Most existing learning platforms that teach Git follow a textbook like approach, teaching users individual commands and explaining their isolated effects. And while some integrate interactivity and visualization to simulate project-like challenges~\cite{learngitbranching}, most platforms completely overlook the collaborative dimension of version control.
A recent intervention GitKit~\cite{braught_2024} aims to addresses this by integrating learning activities, instructor guides and simulated community responses within a snapshot of an open-source project, thereby mimicking the large-scale collaborative aspect of Git. 
Still, authentic coordination requires interacting with partners who hold different mental models of code and Git, which is difficult for individual learners to reproduce alone or by simulating others' actions. 
This necessitates designing learning environments that let Git learners experience how their work intertwines with a partner’s -- developing awareness of collaborators’ actions, perspectives, and contexts.
In short, \emph{mastering Git takes two}.

To address this gap, we introduce \sys{}, a browser-based training platform that leverages collaborative learning pedagogy to enable learners practice Git in realistic, distributed environments. 
\sys{} embeds a full Git instance directly in the browser, allowing users to execute native Git commands inside resettable, sandboxed workspaces. 
Learners can walk through interactive exercises and visualizations of Git’s internal states, with the freedom to reset or reuse workspaces for safe experimentation. 
To resemble real-world distributed workflows, \sys{} provides a collaborative mode where two learners share a remote repository and must coordinate their actions -- echoing the design of \rev{the cooperative two-player video game} It Takes Two \cite{fares2021ittakestwo}.
\rev{Similar to collaborating on Git, this game places both players in the same environment, performing independent actions that need to be coordinated in order to make progress. The game features a persistent split-screen view, with each players' perspectives displayed side by side for shared awareness for the cooperative game play.}
\rev{Likewise,} to support productive collaboration, \sys{} \rev{in its collaborative mode} aims to foster social presence~\cite{lowenthal2010social} by making partners’ local contexts and actions visible. 
It integrates a mirrored view of each peer's screens, streaming commands, file changes, and repository states in real time to enable synchronous reasoning about distributed state. 
Unlike GUI clients or professional IDE integrations, \sys{} is not designed for everyday software development. 
Instead, it serves as a training simulator: a safe yet authentic environment where learners can rehearse Git’s coordination challenges and experience firsthand that effective version control takes two.

In order to investigate the influence our intervention, \sys{}, on learning and practicing Git, the main research question we pose in this paper is :
\begin{itemize}
    \item[] \textit{How does providing contextual awareness of collaborators’ actions during Git tasks affect learners’ conceptual understanding of Git’s distributed nature and shape their collaborative behaviors when using Git?}
\end{itemize}
To answer our research question, we conducted a controlled within-subject study with 26 Git users working in pairs, comparing their coordination and collaboration on distributed Git tasks when using \sys{} with the mirrored view versus a standard single-view setup. 
In both conditions, participants worked with their own local repositories and a shared remote repository; the only difference was whether they could see their collaborator’s mirrored view. 

Our findings indicate that the split-screen interface of \sys{} facilitated peer awareness of the distributed processes underlying Git, and supported peer learning and social presence. Although no significant performance differences were observed between conditions, the split-screen interface significantly reduced users' mental workload. In particular, viewing their partners' local Git tree and terminal streamed in real time facilitated awareness of their collaborators' contexts  and actions, and participants consistently preferred the split-screen interface over the regular single-view baseline.
Beyond the impact on instructional platforms for teaching Git, we find that the results speak positively on the use of split-screen interfaces in collaborative learning contexts, and merit further investigation in educational research and classroom usage, and as well as in other domains of distributed workflow.

%% file: sections/02_related_work.tex
\section{Related Work}\label{sec:related_work}

\subsection{Collaborative Programming and Pair Programming}
In software development and CS education, collaborative learning is a widely adopted pedagogy that is known to foster programming abilities and critical thinking~\cite{gokhale, chowdhury}.
Pair programming~\cite{Beck1999} is a prevalent and effective collaborative programming paradigm, involving two co-located programmers working together at the same computer. Traditionally one acts as the \textit{driver}, writing the code, while the other takes the role of \textit{navigator}, reviewing the code and providing feedback---swapping roles periodically. In recent years with the advent of remote work and online learning this form of collaboration has also extended into remote formats, termed as distributed pair programming.

Prior research has empirically demonstrated the benefits of pair programming for both professional developers and learners in computer science. In professional contexts, it facilitates knowledge sharing, improves attention and productivity, and contributes to the development of higher-quality software products~\cite{SillittiEtAl2012, WilliamsEtAl2000, Zieris2020, ChongEtAl2006}. In educational settings, pair programming has been shown to enhance learning outcomes~\cite{Williams2001}, improve code quality~\cite{McDowell2006, DybåEtAl2007}, and support student retention~\cite{McDowell2002}, with these pedagogical benefits also extending to distributed pair programming~\cite{BigmanEtAl2021, Hanks2005}.

Research has shown that the interactions and role alternation inherent in pair programming help learners organize their knowledge and deepen their conceptual understanding~\cite{WernerEtAl2004}. Moreover, novices who pair program are more confident in their work than those who do not pair~\cite{McDowell2003}. While role-playing in pair programming is often unstructured, effective learning depends on how well \textit{drivers} and \textit{navigators} coordinate their actions and alternate roles. To facilitate this, researchers have integrated mechanisms such as distributed resources to be used by pairs to guide turn-taking. For instance, Pyrus~\cite{ShiEtaAl2019} employs game mechanics to encourage planning and coordination, while Hayatpur et al.~\cite{HayatpurEtAl2023} structured Parsons’ problems into individual and shared synchronized spaces to foster interdependence between pairs. However, other than scripting coordination, another important factor for pairs to learn and practice programming together, especially in remote formats, is to make collaborators aware of each other’s context.

Several tools have been developed to support distributed pair programming by fostering remote awareness between peers. For example, RIPPLE (Remote Interactive Pair-programming Learning Environment)~\cite{BoyerEtAl2008}  supports distributed synchronous collaboration in classrooms through a shared programming view that mirrors peers’ actions. To reduce coordination effort between remote pairs, D’Angelo and Begel~\cite{DAngeloBegel2017} designed a Visual Studio extension that provides gaze-awareness by synchronizing collaborators’ scrolling, highlighting, and window switching. Similarly, PearProgram~\cite{BigmanEtAl2021} supports synchronous collaboration with structured driver–navigator roles, offering a shared IDE that includes code and output windows, real-time annotations, and embedded prompts. \rev{More recently, Colin et al.~\cite{ColinEtAl2024} introduced a tool that provides synchronized editing, shared-document awareness through integrated view-port, concurrency control, and live communication, enabling distributed pair programmers to collaborate as if co-located.}

Although these tools effectively support distributed pair programming and can extend to other forms of collaborative programming, they primarily facilitate coordination and awareness through a single shared code view that maintains a synchronized state. While such a view makes learners’ immediate actions visible and helps teams coordinate low-level code interactions, these actions alone do not provide sufficient context to reveal peers’ reasoning processes. As a result, learners have limited opportunities to maintain awareness of each other’s mental models and in turn to shape and drive their own learning~\cite{GoswamiEtAl_PD2023}. To foster long-term learning through productive collaborative exchanges, it is therefore crucial to integrate scaffolds into synchronous programming workspaces that allow collaborators’ thought processes and actions to be externalized and persistently visible to peers.

\subsection{Collaborative Editing and Shared Awareness}




To enable collaboration on code and other contents, along with research artifacts, there exists several commercial tools. Programming platforms such as \texttt{Studio Live Share} \cite{visualstudioliveShare} for Microsoft's Visual Studio, \texttt{Code With Me} \cite{codeWithMe} in JetBrains IDE, web-based collaborative working environments like \texttt{CoCalc} \cite{cocalc}, Google's \texttt{Colab} \cite{googleColab}, \texttt{JupyterLab} \cite{jupyterlab}, and document editors like \texttt{Google Docs} \cite{googleColab}, \texttt{Overleaf} \cite{overleaf} allow multiple users to work on the same content simultaneously -- each in an independent view-port with collaborator's awareness cues like colored cursors or user labels. By centralizing the actions of different users into a single view, these tools obscure the distributed nature of collaborative processes to stream-line coordination. However, while these smooth coordination supports productivity, simultaneous editing can often lead to conflicts in tasks and code~\cite{YangEtAl2023}.

Prior research in HCI and CSCW has long emphasized the importance of shared awareness of collaborators' activities--both for coordinating efforts in collaborative work\rev{~\cite{DourishEtAl1992, GutwinEtAl1996, GutwinEtAl1998, Dix1994} } and for supporting collaborative learning in educational contexts~\cite{GutwinEtAl1995}. 
This dynamic awareness, which is equally critical in remote collaborative programming and learning, is often conveyed through single synchronized interfaces as mentioned earlier. 
However, studies have shown that on such single-view synchronous code-editing interfaces, collaborators' activities and reasoning are not clearly communicated, leading to unbalanced participation and interference without strategic coordination~\cite{GoldmanEtAl2011, WangEtAl2019}. 
Moreover, while collaborating during programming, an individual's coding and debugging can be an iterative process, involving exploration and trial-and-error. 
This can account for overwhelming amount of actions in a single shared-view interface for collaborators. Therefore, collaborative systems for programming learning should incorporate features that allow individual approaches to unfold while making learners' local context visible to remote peers for mutual awareness\rev{~\cite{GoswamiEtAl_PD2023}}.


Multi-view interfaces have been leveraged across diverse collaborative contexts, both co-located and remote, to enhance awareness and support effective coordination. For example, by integrating a set of multiple views, shared, private, and partially shared views of a table-top display, Permulin~\cite{LissermannEtAl2014} enables collaborators to switch between tightly coupled teamwork and independent work seamlessly. In programming teaching contexts, multi-view interfaces have been used in system like Codeopticon to help tutors monitor multiple students simultaneously and provide interactive support~\cite{Guo2015}. Similarly, PuzzleMe~\cite{WangEtAl2021} offers multi-view interfaces for not only instructors to monitor students but also enable peer collaboration, providing separate tabs for students' own code and their group members' code, enabling them to review and give feedback on each other's work. Another notable study exploring this approach is Cocode~\cite{ByunEtAl2021}, which projects mirrored views of all peers' code activities and outputs from their respective editors into side-by-side multiple panels on a learner's workspace, allowing students to observe multiple collaborators simultaneously. However, Cocode was designed to foster social presence~\cite{lowenthal2010social} asynchronously in large online classrooms, rather than to support real-time, synchronous collaborative programming.

Although multi-view interfaces have the potential to bring simultaneous and consistent awareness of collaborators' contexts, actions, and intentions, while giving opportunity for individual exploration. Yet they have received little attention in design of synchronous collaborative programming learning platforms. 

\subsection{Git Learning and Version Control Education}
\begin{table*}[ht!]
\centering
\caption{Existing Git tutorials and learning platforms}
\label{tab:existingMaterial}
\small
\resizebox{0.9\textwidth}{!}
{%
\begin{tabular}{lp{3cm}p{3cm}p{3cm}}
        \toprule
    \textbf{Platform/Tutorial} & \textbf{Git installation} & \textbf{Multi-User Collaboration} & \textbf{Interactive Git Tree Visualization}\\
         \midrule
         LearnGitBranching \cite{learngitbranching}  & emulated & no & yes\\
         Git Immersion  \cite{gitimmersion}  & local & no & no\\
         gitexercises.fracz.com \cite{gitexercises} & local & no & no\\
         Educative Git Quiz \cite{educativeGit} & local & no & no\\
         Berkley Interactive Git Tutorial \cite{interactiveGitTutorial} & local & no & no \\
         Codecademy Git Tutorial  \cite{codecademy} & web-based, sandboxed & no & no\\
         Git How To \cite{gitHowTo} & local (VS Code extension) & no & no\\
         Try Git  \cite{tryGit} & local & no & no\\
         \midrule
         \textbf{GitAcademy} & web-based, sandboxed & yes & yes\\
         \bottomrule
\end{tabular}%
}
\end{table*}
Learning Git for software development and programming has become increasingly relevant, yet research highlights several challenges new learners, and as well as developers face when learning and using Git~~\cite{rosso_jackson_2013, Church2014, Isomottonen2014}. Git's extensive list of commands that modify its states and high degree of interactivity among its components poses significant cognitive load for learners and reduces its overall learnability, especially for beginners~\cite{Hynes2015}. Studies have found that beginners frequently struggle with both conceptual understanding and practical application of Git -- like applying developer workflows, resolving merge conflicts, and distinguishing between local and remote repositories, as well as often misuse commands or avoiding advanced features altogether~\cite{Isomottonen2014, eraslan_2020}. Furthermore, mastery of Git is not limited to its commands alone; it also requires the ability to reason about distributed states and components in Git and to coordinate actions across collaborators in alignment with Git’s conceptual model~\cite{rosso_jackson_2013}. 

Previous research has explored various methods and tools to aid learners processes of learning Git. Students are often exposed to learning Git in standalone CS classrooms or in group-project contexts~\cite{missingsemester2020, tushev2020using}. Lightweight tutorials have been shown to encourage students to adopt more advanced Git workflows~\cite{sproull}, while integrating GitHub into coursework provides insights into collaboration patterns and underscores the importance of structured guidance~\cite{valerio, niki}. Using Git itself as a course management tool reinforces usage through continuous exposure~\cite{haaranen}, and combining Git-based workflows with ongoing assessment can enhance learning and engagement with version control~\cite{gary}. These pedagogical approaches build learners’ familiarity with Git in tightly scoped classrooms, yet in practice most acquire Git commands and concepts on demand~\cite{yang2022}, within real-world project contexts. This underscores the need for interactivity in pedagogical materials and hands-on approaches to help students overcome both conceptual and practical challenges.

In this regard, Chen et al.~\cite{chen} introduced a Git Education Game that teaches core Git concepts through game mechanics, boosting motivation and engagement while allowing students to safely experiment with commands and learn abstract topics like branching, merging, and conflict resolution through trial and error. Similarly, hands-on Git training with lectures, videos, and interactive labs has been shown to improve student confidence and highlights the importance of practice and feedback~\cite{elgun}. Although, research highlights the importance of realistic project-based courses to help students recognize the value of version control systems and better understand Git's collaborative use~\cite{Isomottonen2014}, these pedagogics often overlook explicitly teaching the collaborative dimension of using Git which is inherent to its distributed nature.

The existing landscape of Git learning material is extensive. Many platforms \cite{gitimmersion, gitexercises, educativeGit, interactiveGitTutorial, codecademy, gitHowTo, tryGit} follow a textbook style of teaching Git. Content is structured to provide a high level understanding of Gits mental model, followed by a series of hands-on exercises that learners can execute on their local machines, and see the effects of Git commands in isolation. Two platforms notably distinguish themselves from the rest: LearnGitBranching \cite{learngitbranching} and Codecademy \cite{codecademy}. LearnGitBranching teaches Gits mental model by interactively visualizing the Git tree. Codecademy on the other hand is unique as it is the only platform which exposes a real Git instance to web-users. This lets them leverage sandboxed environments that can execute the full range of Gits functionality. An overview of the existing Git learning platforms are summarized in Table~\ref{tab:existingMaterial}.

Although these platforms simulate realistic interactions with Git, they do not embody and expose its distributed collaborative nature. Only a recent intervention, GitKit~\cite{braught_2024} addresses this challenge  by combining learning activities, guided instruction, and simulated community feedback within an open-source project snapshot, providing learners with a realistic collaborative Git experience. However in real-life, learners need to interact with others when working with Git, which requires them to understand their collaborator's mental models of code and Git -- which is difficult for individual learners to reproduce alone or through simulation. \rev{Ma et al.~\cite{MaEtAl2022} further highlight that developers benefit from immediate awareness of collaborators’ edits and Git states, suggesting that real-time awareness tools could support Git learning by making these otherwise invisible version-control processes observable.}

To address the challenges of providing realistic and collaborative Git learning experiences, we extend the concept of distributed pair programming. Our approach aims to foster Git learning by integrating synchronous collaborative learning scaffolds that convey and train on Git's inherent distributed nature. Specifically, this scaffold introduces a real Git distributed workflow where progress depends on joint action by collaborators, inspired by the design of cooperative game, It Takes Two \cite{fares2021ittakestwo}. To make pairs' context visible to each other, we leverage a dual-view, split-screen interface: providing individual autonomy and maintaining shared awareness through a mirrored view of the peer's environment. We present our intervention \sys{}, which combines interactive visualization and hands-on real Git environment, while explicitly training learners in authentic coordination and collaborative practices essential for real-world Git usage.

%% file: sections/03_system_design.tex
\section{System Design} \label{system-design}
We present \sys{}, a web-based learning platform designed to address the challenges outlined in Section \ref{sec:related_work}. 
\sys{} offers a real Git environment exposed through the web, enabling users to run native Git commands directly from their browsers. 
The platform utilizes sandboxed docker containers to ensure each user has a safe and isolated workspace, which can be reset or replicated as needed for repeated practice.

To make the system more concrete, we illustrate its features through the perspective of two learners, Cody and May, who are practicing the branching and merging exercise together, as shown in Figure \ref{fig:teaser}.
Their scenario demonstrates how core features of \sys{} support Git's distributed learning challenges in practice.

\subsection{Design Goals}~\label{design_goals}
\sys{} was designed to foreground Git’s inherently distributed and collaborative nature. 
\rev{Our design decisions are informed by prior research on the effective design of educational groupware technologies~\cite{GutwinEtAl1995, GutwinEtAl1996, GutwinEtAl1998}. According to Gutwin et al., \textit{workspace awareness}, defined as an up-to-the-moment understanding of another person's interactions within a shared space~\cite{GutwinEtAl1996} is critical for supporting productive collaborative learning in remote settings.  Workspace awareness facilitates the coordination of tasks and resources and helps learners transition between individual and shared activities~\cite{GutwinEtAl1995, GutwinEtAl1996}. Learners can use their knowledge of peer's workspace to interpret peer's actions, anticipate their next steps, and reduce the effort needed to coordinate shared tasks and resources~\cite{GutwinEtAl1996}. It also creates opportunities for peer learning, such as shadowing peer's skills, offering assistance, resolving differences in understanding, and building shared meaning~\cite{GutwinEtAl1995}.
When working with Git, these needs become especially important. Understanding the distributed activities of collaborators is essential for coordinating work and navigating challenges that emerge when multiple users share a repository. Hence, rather than centering solely on individual workflows, \sys{} highlights the awareness and coordination demands inherent in collaborative Git use. }
This focus led to four key design goals.

\subsubsection{Full Functionality for Collaboration}
Learners should experience the complete workflow of distributed version control without the overhead of manual setup and configuration. \rev{However, they often struggle to perceive the collaborative nature when working individually with Git. This design goal aims to make Git’s distributed nature explicit by allowing learners specifically to experience the collaborative aspects of working with Git in a distributed setup. To achieve this,} \sys{} automatically provisions a collaborative environment consisting of a shared remote repository and two connected local repositories \rev{for collaborative projects}, eliminating the need for learners to configure remotes, clone peers’ repositories, or manually seed divergent histories. \rev{By offering a pre-configured collaborative workspace, \sys{} allows} practice sessions to begin immediately with realistic coordination scenarios such as merging or resolving conflicts.a

\subsubsection{Distributed-State Visualization}
\rev{This design goal seeks to externalize Git's distributed-state reasoning through visual affordances of collaborators' evolving local repositories along with their shared remote repository.} 
\sys{} presents both the learner's \rev{own} local repository and their partner's \rev{local} repository in parallel, updating dynamically as actions occur.
By making the evolving states of both sides visible to users, the interface supports a conceptual understanding of how operations propagate across distributed histories. \rev{This feedthrough~\cite{Dix1994}, i.e, the observable effect of peer's action on workspace artifacts, like local and shared repositories, provides workspace awareness~\cite{GutwinEtAl1996} in our distributed system. This intends to help learners grasp how independent work diverges, synchronizes, and integrates across collaborators in Git.}


\subsubsection{Awareness of Peer's Workspace}
Git learning requires not only understanding one’s own commands but also anticipating collaborators’ actions. 
\rev{In face-to-face settings, learners can use gestural references or deixis, such as pointing to code in each other's systems, to make references clear, easing coordination and explanation. However, in remote environments, such gestural deixis is largely absent, and language alone is often insufficient for referring precisely to states or events within a distributed workflow~\cite{BrownEtAl1989, Dix1994, GutwinEtAl1995}. Hence, this design goal aims to provide a comprehensive awareness of the peer’s remote workspace through the use of deictic features.} \sys{} integrates a mirrored interface that allows each learner to observe their partner’s terminal inputs, file edits, and commit history in real time. \rev{This awareness layer is intended to facilitate explanations, support peer teaching, and enable learners to reason about coordination problems, such as divergent histories or conflicting commits, that cannot be understood or reproduced alone.}

\subsubsection{Easy to Access and Resettable Setup}
To lower the barrier to entry, the system runs entirely in a web browser with no local installation. 
Each environment is sandboxed in containers that can be reset or replayed, providing a safe space for experimentation. 
Learners can explore high-stakes Git actions, such as rebases or force pushes, without risk to their real projects.

Together, these goals frame \sys{} not as an everyday development tool, but as a training simulator where learners can safely and authentically rehearse the coordination challenges of distributed version control.

\begin{figure*}[!ht]
    \centering
    \includegraphics[width=0.9\textwidth]{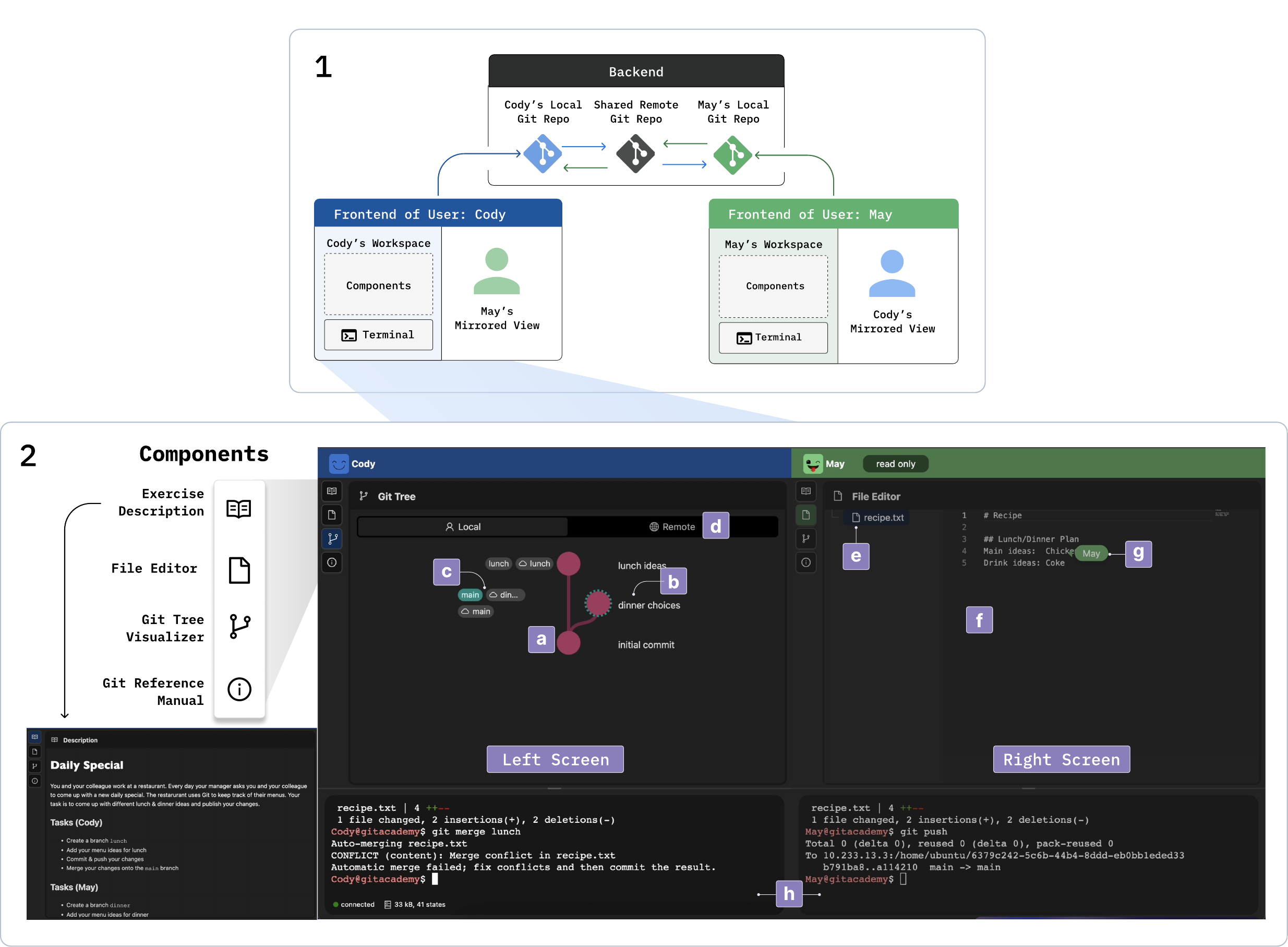}
    \Description{Two-part figure showing the GitAcademy split-view system. The top diagram depicts two users, Cody and May, each working in separate front-end environments connected to individual local Git repositories and a shared remote repository. Each front-end displays the user’s own workspace alongside a mirrored, read-only view of the peer’s workspace. The bottom diagram shows Cody’s split-view interface in detail, with a Git tree visualizer and terminal in the left panel and a read-only view of May’s file editor activity in the right panel, alongside a vertical menu for exercise instructions, file editing, Git visualization, and reference materials.}
    \caption{ \rev{\sys{} Design Overview: 1. Implementation of the split view mode. 2. Split screen interface from the Cody's perspective.
    \textbf{Left screen}: Cody's own workspace shows a Git tree visualization of his local Git repository, with (a) commits visualized as round bubbles, (b) commit messages, (c) branch references and (d) tab bar to switch between local and remote repository Git trees.
    \textbf{Right screen}: May's workspace and actions, mirrored and read only to Cody. In current view May is on File Editor component with (e) a list of files in the exercise directory, (f) a web file-editor and (g) mouse-pointer of May on her own screen (as seen by Cody) and (h) terminals.}}
    \label{fig:overview}
\end{figure*}

\subsection{Starting an Exercise}

To begin a session, learners access \sys{} through their web browser and select a username and avatar for themselves. 
From the dashboard, a learner can select an exercise and open a new practice room. 
The system generates a unique invitation code, which a partner can use to join the room. 
Once both learners are present, \sys{} provisions the collaborative environment: two sandboxed local repositories connected to a shared remote repository. 
The pair can then immediately start practicing, without any additional setup or configuration.

\subsubsection{Example:} Cody chooses the \say{Branching and Merging} exercise and opens a new room. 
He shares the invitation code with May, who joins from her own laptop. 
In seconds, the system sets up their distributed environment. 
Both learners see their personal local repository linked to the same remote, ready for collaborative practice.

\subsubsection{Implementation:} When Cody and May create a new exercise room, the backend spawns two Docker containers for local repositories and one container for the shared remote. Cody's and May's frontend then communicates with an environment manager in the backend which provides them with up-to date git state information and executes all their terminal commands in their respective local repositories.
To create meaningful coordination scenarios, each exercise includes a setup script that modifies the initial repository states: for instance, by adding an extra commit to May’s local branch while leaving Cody’s unchanged. 
This ensures that learners immediately encounter distributed states that require push, pull, or merge operations, rather than starting from identical clones.

\subsection{The Exercise Room}
\rev{As shown in Figure \ref{fig:overview}\rev{.2}, when two learners join a practice room, the interface presents a coordinated set of components designed to scaffold Git learning: an \textbf{exercise description} orients learners to the task and sets the collaborative focus; a \textbf{file editor} (Figure \ref{fig:overview}\rev{.2}(f)) supports hands-on code modification, helping learners connect repository changes to their underlying source; a \textbf{Git tree visualizer} (Figure \ref{fig:overview}\rev{.2}(a), (b),(c)) provides an interactive view of both local and remote repository histories. Learners can click on the tree nodes to inspect commits and references, such as commit hashes, authorship and timestamps, therefore build awareness of the evolving shared state; a \textbf{reference manual} (Figure \ref{fig:screenshot_split_reference}) offers searchable Git documentation, encouraging just-in-time conceptual clarification; finally, the \textbf{terminal} (Figure \ref{fig:overview}\rev{.2(h)}) allows learners to execute native Git commands.}
All components are designed to work together and automatically react to changes in the repository state. Should git change the contents of a file, a user add a new commit or create a new branch, all changes are  tracked and updated in the UI components.



\subsubsection{Example:} After entering the room, Cody and May review the exercise description, which asks them to create diverging branches and later merge them. 
Cody edits a source file in the built-in file editor and commits his changes through the terminal. 
Immediately, both learners see the new commit appear in the Git tree visualizer, along with the diverging branch histories. 
When May is unsure about the correct rebase syntax, she opens the reference manual directly within the interface, rather than leaving the exercise environment.

\subsubsection{Implementation:} Each component is implemented as a modular React widget. 
The file editor syncs with the underlying Docker container’s working directory, ensuring that changes reflect exactly as they would in a local repository. 
The Git tree visualizer queries the Git process to render commit graphs, showing both local and remote states. 
The reference manual wraps Git’s native main pages in a searchable interface. 
The terminal is streamed directly from the container, allowing full Git command execution with real-time feedback. 

\subsection{Collaborative Features}\label{system:collab_features}

\begin{figure*}[tp!]
    \centering
    \includegraphics[width=0.7\linewidth]{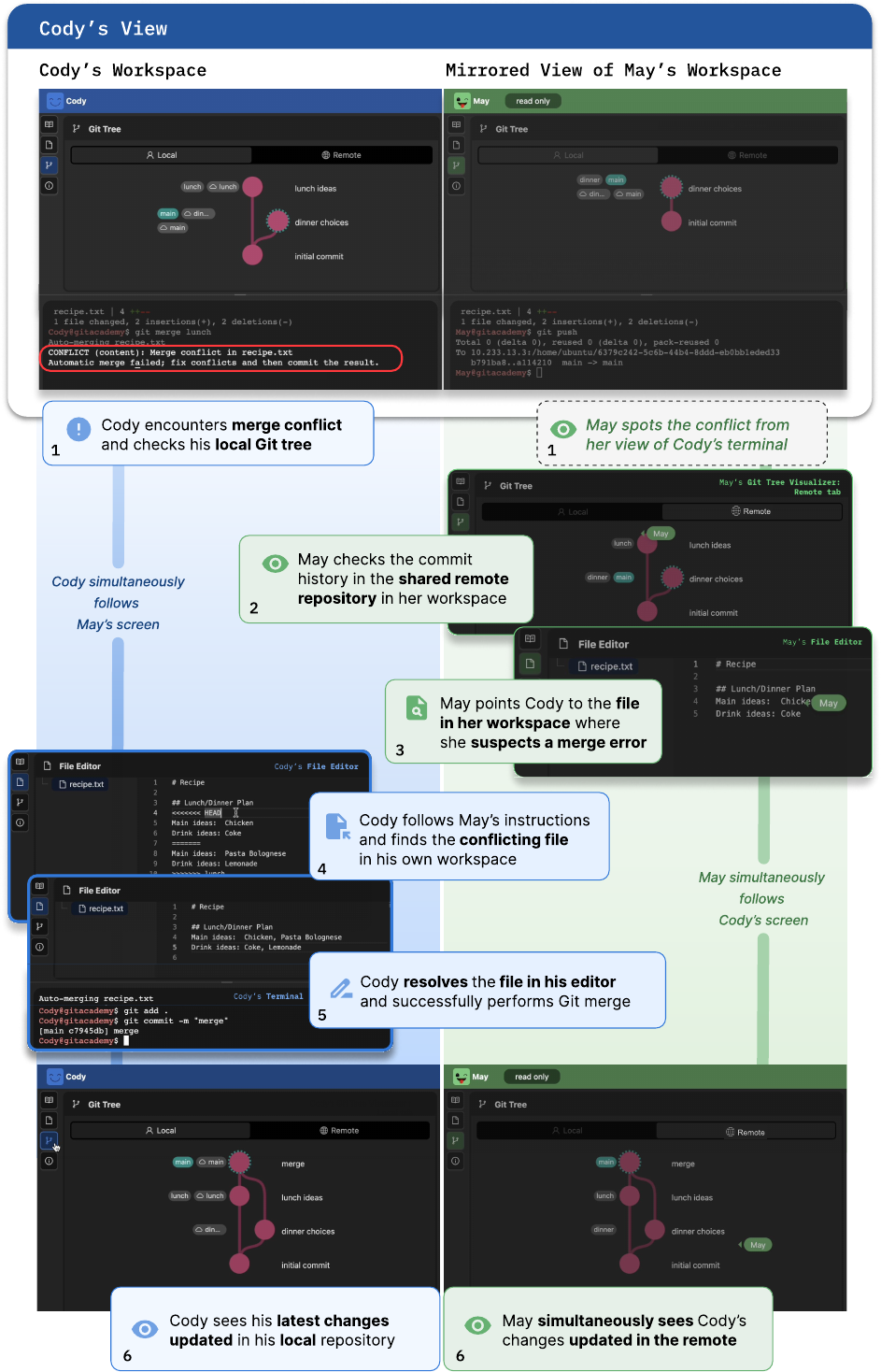}
    \Description{Diagram of the GitAcademy interface showing a distributed Git workflow from Cody’s point of view. Cody’s workspace appears on the left and a mirrored, read-only view of May’s workspace appears on the right, allowing each user to see the other’s actions. The interface shows a merge conflict in Cody’s terminal, Git tree visualizations, and numbered description boxes indicating simultaneous actions, with dashed boxes representing implicit actions inferred by May through awareness of Cody’s workspace.}
     \caption{\rev{Example of a distributed workflow in \sys{} as experienced in Cody’s front-end. Both Cody and May have mirrored views of each other’s workspaces, allowing them to follow each other's actions. Cody encounters a merge conflict, and May assists him using collaborative features available to both. Dashed description boxes represent implicit actions by May based on her awareness of Cody’s workspace. Description boxes with the same number indicate actions occurring simultaneously.}}
    \label{fig:workflow}
\end{figure*}

To support \rev{comprehensive awareness of} distributed workflows, \sys{} extends the exercise room with features that make a partner's actions visible while preserving individual control. \rev{Prior work on distributed pair-programming tools has explored two key capabilities for workspace awareness: providing a shared view of the same artifact, such as screen sharing, and supporting distributed coordination~\cite{WinkerEtAl2010, ColinEtAl2024}.} \rev{In line with this,} \sys{} offers a \textbf{split-view interface} that displays both learners' complete environments side by side providing real-time update of peers' workspaces to each other. \rev{This mirrored view serves as the foundation for workspace awareness~\cite{GutwinEtAl1995} in our system and provides learners a persistent, contextual sense of how their partner's actions relate to their own. Following the proximity compatibility principle~\cite{WickensEtAl1995}, placing the two workspaces close together aimed to reduce cognitive effort on task by aligning the mental operations required to interpret code edits, commits, and terminal commands. Additionally, this mirrored view also aims to foster learners' social presence for peer-learning~\cite{ByunEtAl2021}.}

\rev{To further enrich workspace awareness, \sys{} adds \textbf{deictic cues} across the mirrored views: \textbf{a shared cursor} indicating where a partner is editing or in their workspace and \textbf{real-time terminal mirroring} showing their Git commands. These cues help learners interpret intent, reduce ambiguity during coordination, and support observational learning, allowing learners to pick up practical Git strategies by watching peers~\cite{Kehrwald2008}. The mirrored view also lets \textbf{learners see when their partner is interacting with other components}, like inspecting their local Git tree, which is particularly useful for collaborative Git conflict resolution. 
Together, these features intend to foster peer-learning, and turn the mirrored workspace into a coordination scaffold that keeps collaborators aligned as their shared repository evolves.}

\subsubsection{Example:}\label{distributed_workflow} During a merge task, Cody attempts to integrate May's branch but encounters a conflict (see Figure \ref{fig:workflow}).
In the split view, May immediately sees Cody's error message and inspects the Git tree visualizer to diagnose the issue.
Using the shared cursor, she points to the conflicted lines in Cody's file and talks to him through the resolution steps.
Cody follows her instructions and edits the file on his own screen.
Importantly, May can not directly take control of Cody's workspace; instead, \textbf{the system encourages guidance and explanation, reinforcing learning through collaboration rather than substitution}.
On the other hand, if May had been less familiar with merging, the mirrored terminal and Git tree visualizer would still allow her to follow Cody's process step by step, ask clarifying questions, and build her own understanding from observing his actions.
This design supports both asymmetric peer teaching and mutual exploration, so either learner can take the lead depending on the task and their level of confidence.

\subsubsection{Implementation:} Split-view rendering (Figure~\ref{fig:overview}.1) is achieved by duplicating the component layout of each learner's room.
In addition, users' actions such as cursor movement and keyboard inputs are 
collected separately in both frontends and streamed in real time to an environment manager running in the backend. The environment manager keeps track which users are present in a room and broadcasts incoming messages accordingly. It also takes care of executing terminal commands on the underlying docker containers.




%% file: sections/05_evaluation.tex
\section{User Study}
To evaluate how \sys{}’s split-view interface influences collaboration between pairs of Git learners, we conducted a randomized controlled user study. The study was approved by our institution’s IRB and was pre-registered\footnote{See pre-registration: \url{https://aspredicted.org/m2k5-bs5k.pdf}.} 
Our primary goal was to investigate whether providing a shared visual context of a remote peer’s local environment through a split-screen interface supports productive collaborative experiences of practicing Git and enhances learning outcomes. Accordingly, we pose the following research question: 
\rev{\textbf{RQ. How does \sys{}’s split-view interface shape learners’ behaviors, experiences and performance in collaborative Git practicing?}}

\rev{As outlined in our design goals (Sec.~\ref {design_goals}), \sys{}'s split-screen interface is intended to foster workspace awareness of peers’ activities, help learners interpret distributed states for coordination in Git, and create opportunities for peer teaching and learning. To investigate how these intended affordances unfurl collaboration in practice for Git learning, our study aimed to: (1) establish the relevance of collaborative practice for Git performance, (2) evaluate the influence of \sys{}’s split-view interface on exercise performance and learners’ cognitive load, (3) examine whether and how the split-view interface supported productive Git collaboration behaviors, and (4) identify which features in the \sys{} learners found most supportive for Git collaboration.}

\frev{Accordingly to investigate its influence on Git performance, we evaluated the following hypothesis: \textit{H1. Participants will demonstrate higher comprehension of Git concepts when using the split-screen interface for collaborative learning of Git compared to the traditional single-screen interface.} 
We also conducted further exploratory quantitative and qualitative analyses to examine how the split-screen interface shaped peer-awareness and collaborative Git learning behaviors, and how these processes were related to performance outcomes.}

\subsection{Study Conditions}
Our study employed a within-subjects design with two conditions: an experimental condition and a baseline condition. Participants were arranged into pairs, and each pair were subjected to both conditions. In the experimental condition, pairs used the split-view interface of the \sys{}, which displayed their partner's screen alongside their own workspace. In the baseline condition, pairs used a regular single-screen interface, where only their own workspace was visible. In both conditions, participants had to solve Git exercises collaboratively and were able to communicate through Zoom. Participants were paired according to their reported availabilities and the time of sign-up. 

\subsection{Study Procedure}
\begin{figure*}[!htb]
    \centering
    \includegraphics[width=\textwidth]{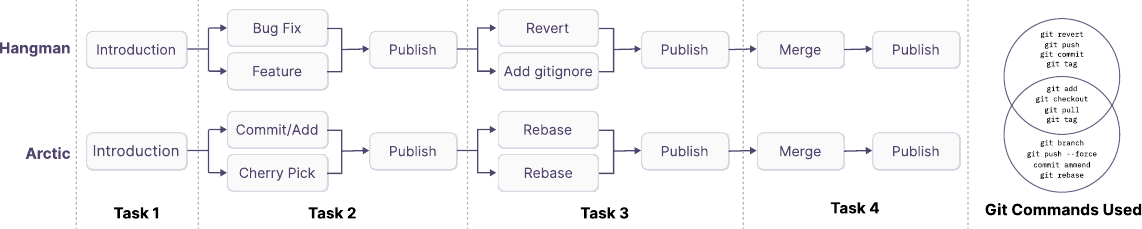}
    \Description{Diagram showing two parallel, four-step exercise workflows labeled Hangman (top) and Arctic (bottom), progressing from left to right across Tasks 1 to 4. Each workflow begins with an introduction, branches into different sub-tasks in Tasks 2 and 3, and concludes with a merge and publish step in Task 4. A separate panel on the right lists Git commands used across the exercises, with overlapping commands indicated between the two workflows.}
    \caption{\rev{Exercises.} The exercises structure of \rev{\textit{Hangman} and \textit{Arctic}. Each consist of different tasks that follow the same exercise outline.} Tasks 2 and 3 requires participants to split up and perform different sub-tasks. \rev{The exercise require same number of Git commands with a set of operations.}}
    \label{fig:exercise}
\end{figure*}

Below we describe our study procedure in details. The average duration of each study session was \mean{105.4} minutes (\sd{8.1} minutes). 

Prior to the start of our study, participants first completed a demographic survey (Table~\ref{table:demo_measures} in Appendix~\ref{appendix}) at home in which they provided background information, reported their familiarity with various Git commands, and described how frequently they used Git in both personal and collaborative settings. 

On the day of the study, pairs of participants met in-person on the university campus. Each session began with a introduction of the study and signing of the consent form by the participants. At the beginning of the study session, each person from the paired group had to individually complete a warm-up introductory task to familiarize themselves with the \sys{} platform and the study setup. This warm-up ensured that participants understood the interface and were prepared for the study.  
Following this, each member of the pair had to individually complete a pre-test Git proficiency assessment, described in Sec.~\ref{sec:pre-git_test}. This test evaluated their understanding of Git commands and concepts, offering a measure of their command-line Git expertise. 

After completing the pre-test assessment, participant groups were asked to work on one collaborative Git exercise in each condition of the study. Two different exercises, \textit{Arctic} and \textit{Hangman}, were designed for this purpose, which are described in Sec.~\ref {sec:task_materials}. 
Participants completed each exercise under a different study condition. \rev{The order of exercises and their assignment to study conditions were counterbalanced across pairs to mitigate potential transfer of learning or order effects (Table~\ref{table:counterbalance} in Appendix). We used a 2×2 counterbalanced design~\footnote{The python script used for counterbalancing provided in the supplementary materials.} combining task order (\textit{Hangman}$\rightarrow$\textit{Arctic} vs. \textit{Arctic}$\rightarrow$\textit{Hangman}) and study condition (regular vs. split-view), ensuring each of the four combinations was assigned equally across the signed-up groups. The final pair, who enrolled after the counterbalancing was completed, received a randomly assigned condition.}

Before beginning the exercises, each participant pair was separated into two rooms, each supervised by one of the authors of this paper. 
Participants joined a Zoom session to communicate verbally with their remote partner. They then worked on the assigned exercise using the \sys{} interface corresponding to their study condition. 
While solving the exercises, participants were free to speak with each other verbally and could turn on their video if desired; however, they were not permitted to share their screens, in order to maintain control over the study setup. An exercise consisted of four tasks, which were handed out progressively by the researcher in respective rooms, once the pair agreed with each other and confirmed they had completed the current task or chose to move on to the next. 
Each exercise session had a 25-minute time limit, after which the pairs were required to stop, even if the exercise was not fully completed. 
At the end of each collaborative exercise, participants completed an individual post-survey assessing workload, task difficulty, collaboration experience, and the level of support provided by the interface. 
After completing the first condition, participants were given a five-minute break. 

Finally, after completing both conditions, participants were interviewed in-person by the accompanying researchers. The interview adopted a semi-structured format and lasted an average of \mean{9.27} minutes (\sd{2.52} minutes). During the study each participant’s computer screen and audio were recorded for later analysis. 


\subsection{Participants}
We recruited a total of 26 participants, forming 13 pairs for the study. This sample size deviated slightly from our pre-registration, \rev{due to two late sign-ups, after the counter-balancing was completed}. Participants were recruited via posters on campus. The inclusion criteria for taking part in the study required them to be above the age of 18. 
All participants gave informed consent prior to taking part in the study. All the collected data was anonymised to ensure privacy, and participants were reminded of their right to withdraw from the study at any time without consequence. 

Of the 26 participants enrolled in the study, eight were female, 17 were male, and one preferred not to disclose their gender. Fourteen participants ranged from 18 to 24 years, and 12 were in the age range between 25 to 34 years.
At the time the study, one of our participants held a high school degree, four people held a bachelor's degree, 21 participants held a master's degree. The participants reported varied levels of Git usage. Ten participants reported using Git daily, eleven reported weekly use, two reported monthly use, and three reported using it rarely. 
Some participant pairs knew each other before conducting our study.
Furthermore all participants reported experience with basic commands in Git such as pushing/pulling to/from remote, creating commits and listing branches. The 90-minute study compensated participants with 31 USD. Detailed information regarding individual participant profiles is provided in Appendix~\ref{appendix}.

\subsection{Exercise and Materials} \label{sec:task_materials}
We designed two contrasting exercises named \emph{Hangman} and \emph{Arctic} \rev{to emulate the distributed nature of real-world Git workflows. The exercises were intended to expose participants to different categories of Git commands and as well the authentic challenges that learners typically encounter. Their design was informed by documented difficulties in learning Git: challenges with staging, committing, understanding the relationship between the working directory, staging area, and repository, branching and merging, and synchronizing with remote repositories~\cite{Isomottonen2014}.} 

%
In the \emph{Hangman} exercise, participants take on the role of indie game developers \rev{collaborating} on a same branch. \rev{Since all work occurred on a single branch, this exercise required participants to coordinate closely when performing basic commands such as \texttt{add}, \texttt{commit}, and \texttt{push}; otherwise conflicts were more likely to occur in this single-branch workflow without the isolated feature branches.}
In contrast, the \emph{Arctic} exercise places participants in the role of arctic researchers who use Git to keep a history of collected ice core samples \rev{in separate branches. In this exercise, participants had to use fundamental commands such as \texttt{add} and \texttt{commit}, and were then required to synchronize work across different branches using more advanced commands like \texttt{merge} or \texttt{rebase}.}
As shown in Figure~\ref{fig:exercise}, both exercises follow the same overall structure and consists of four tasks. Task 1 serves a warm up and is straightforward to complete. Tasks 2 and 3 require the participant pair to work on separate sub-tasks, which must subsequently be synchronized by pushing their respective changes to a shared remote repository. Task 4 can either be completed by a single participant or distributed between both participants, depending on the group’s approach. 
\rev{Together, the two exercises provided complementary Git workflows: one centered on immediate coordination within a single shared branch, and the other on managing parallel work across multiple branches, with coordination emerging later. Although the scenarios differed, both exercises required the same number of commands and relied on a shared set of core Git operations, ensuring that the task designs covered a broad range of coordination demands and Git commands.}


\begin{table*}[!ht]
    \centering
    \small
    \caption{Custom Collaboration Coding Framework. Exercise transcripts are coded into Introduction, Organization, Strategy, Learning Effect (LE), Teaching Peer (TP) and Following Peer (FP) categories.}
    \label{tab:custom_coding_framework}
    \resizebox{\textwidth}{!}
    {%
    \begin{tabular}{llp{5cm}p{7cm}}
        \toprule
         \textbf{Category} & \textbf{Abbrev.} & \textbf{Explanation} & \textbf{Example}\\
         \midrule
         \textit{Organizing} & \textsc{ORG} & Study participants split up their work or talk about the logistics of the exercise. &
        \textbf{P09}: \say{All right, <P10>, which one do you want to do?}
         \newline
         \textbf{P10}: \say{Wait, I still need to read them. User B}
         \newline
         \textbf{P09}: \say{All right, I'll do A.}\\

         \midrule
         
         \textit{Strategizing} & \textsc{STR} & Study participants discuss how to attempt their task. & 
         \textbf{P08}: \say{What did you want to do?}
         \newline
         \textbf{P07}: \say{Interactive rebase.}
         \newline
         \textbf{P08}: \say{Yes}
         \newline
         \textbf{P07}: \say{But yeah, I am not sure that it is actually the solution. It's just an idea.}\\

         \midrule
         
         \textit{Learning Exchange} & \textsc{LE} & Collaborative by nature. Study participants figure out an aspect of Git together. &
         \textbf{P01}: \say{So I'm just gonna try, git checkout}
         \newline
         \textbf{P02}: \say{Yeah, you write, git checkout, I guess. And then Villum (...)}
         \newline
         \textbf{P01}: \say{Its working!}\\
         \midrule
         
         \textit{Teaching Peer} & \textsc{TP} & Explicit by nature. One participant explains an aspect of Git to the other. The learner understands why the proposed solution works. & 

         \textbf{P26}: \say{How do you do it again?}
         \newline
         \textbf{P25}: \say{You can do git rebase and then um dash i.}
         \newline
         \textbf{P26}: \say{The interactive rebase?}
         \newline
         \textbf{P25}: \say{uh yeah i usually tell them also how many commits}
         \newline
         \textbf{P26}: \say{oh yes that makes sense}\\
         \midrule
          
         \textit{Following Peer} & \textsc{FP} & Implicit by nature. A participant visually or verbally follows the instructions of their peer. &
         <\textbf{P23} typing the exact command in the split sreen interface for \textbf{P24} to copy it.>
         \newline
         \textbf{P23}: \say{I would just do this, like git checkout development.}\\
         \bottomrule
    \end{tabular}
    }%
\end{table*}

\subsection{Measurements}
\subsubsection{Pre Test Git Knowledge Measures} \label{sec:pre-git_test}
Before conducting any collaborative exercises in the study, all participants were asked to complete a pre-test Git proficiency assessment. This assessment was conducted on \sys{} and consisted of nine questions that could each be solved by entering a single git command. This included questions like \say{Merge branches X and Y} or \say{Stash all your changes} (see all the questions in Table.~\ref{table:git_assessment} in Appendix~\ref{appendix}). The objective was to have an empirical measure of a participants familiarity with git and for them to recall this knowledge under a time constraint of 15 minutes. Each participant’s final score was calculated as the percentage of correctly answered questions, which we denote as the \textit{pre-test score}. To represent a group’s prior experience with Git, we computed the average of both members’ pre-test scores, which we refer to as the \textit{Pre-Test Group Score}.

\subsubsection{Exercise Performance}
For the quantitative analysis, we calculated each group’s completion score for the assigned exercise in each condition to denote their \textit{performance} measures.
As described in Figure~\ref{fig:exercise}, each exercise consisted of four tasks. 
The completion score was calculated as the percentage of correctly solved tasks by the pair, excluding the results of Task 1. Task 1 served as a warm up task in both the exercises and was intended not to assess participants but to familiarize them with the exercise formats. 
The remaining three tasks are considered to be correctly solved if the local and remote repository states of both participants are equivalent to the one derived in our reference solution. This requires both participants to successfully solve their individual sub-tasks and push all changes to the remote. For each group, the final\textit{ exercise performance} was calculated as the average of each task's binary outcome across the three tasks and expressed as a percentage.

For example, consider two participants working on the Hangman exercise. In Task 2, Participant A implemented a bug fix and successfully pushed the change, while Participant B implemented a feature locally but could not resolve diverging branches when pushing to the remote, even with help from Participant A. Although both feature implementations are correct individually, the task would still be scored as zero because the pair failed to integrate their work into a consistent shared repository state.

\subsubsection{Collaborative Behavior Predictors} 
To understand each group's collaborative dynamics during the Git exercises, the authors first familiarized themselves with the screen recordings and audio transcripts. This review revealed that analyzing the data required attention not only to dialogue but also to participants’ actions within the coding environment. Consequently, we employed Process Coding~\cite[p.96]{Saldana2013} to qualitatively analyze learners’ behaviors, rather than adopting the ICAP framework~\cite{Chi02102014}, which is primarily suited for dialogue analysis. This represents a deviation from our pre-registration. 
Two of the authors together first coded random four study sessions belonging to either study conditions using Process Coding and identified five recurring patterns of behavior. One author then coded the remaining sessions based on these patterns, after which the other author reviewed the codes. We describe the collaborative behaviors along with examples in the Table~\ref{tab:custom_coding_framework}. 
Following this, we computed the frequency of each identified category for every group, which we denote as the group's collaborative behavior predictor.

\subsubsection{Post Test Surveys}\label{method:post-test_surveys} To assess participants' perceived cognitive workload while using the \sys{} in respective conditions, we administered the NASA TLX questionnaire \cite{nasa}, which consisted of six items each rated on a 7-point Likert scale. \rev{This allows us to assess how the split-screen interface's mirrored-view influenced cognitive load during Git tasks, as our design intended to reduce cognitive effort (see Sec.~\ref{system:collab_features}).}
To assess participants' self-reported perceptions of social presence under different \sys{} conditions, we used the same survey employed by Byun et al.~\cite{ByunEtAl2021}, which is based on five dimensions of social presence identified by Kehrwald~\cite{Kehrwald2008}. \rev{This measure captured how aware participants felt of their collaborators. It assessed the perceived reality of the collaborator’s presence and the visibility of their context, emotions, personalities, and personal histories---though \sys{} does not convey learners' personalities or personal histories by design.} 
This survey consisted of five items, each also measured on a 7-point Likert scale. In addition to understand perceived collaborative dimension we also included one additional question: \say{I felt like we were helping each other to accomplish the task}, rated in the exact same way as the social presence survey items.

\subsubsection{Post Test Interview} At the end of the study, we conducted semi-structured interviews with each participant. The questions were designed to gain an understanding of the participants' experiences with the two versions of the \sys{} system and to capture their perceptions of collaborative learning and practicing Git.

\subsection{Data Analysis}
We employed both quantitative and qualitative methods to analyze our data. Below, we outline the statistical tests and qualitative analyses used to address our research question.
\subsubsection{Collaborative Behavior Analysis.}\label{sec:analysis_collab_behavior}
To investigate the role of collaboration in learning Git, we analyzed each group's collaborative behavior to predict their performance on Git tasks.
\rev{We fitted a multiple linear regression model using the \texttt{lm} function in R and applied a stepwise model selection procedure based on the Akaike Information Criterion (AIC) to identify the best-fitting model.}

\rev{
\begin{align*}
\operatorname{lm}(
\text{\gitpushsolved{}} \sim\;&
\text{Pre Test Average}
 + \text{Exercise}\\
&+ \text{Condition}
 + \text{ORG}
 + \text{STR} \\
&+ \text{LE}
 + \text{TP}
 + \text{FP}
)
\end{align*}
}


\textbf{Exercise Performance} was the dependent variable. \rev{The outcome variables included:} the pre-test average of each group (\textbf{Pre Test Average}), \rev{task (\textbf{Exercise}: Hangman vs Arctic)}, condition (\textbf{Condition}: Split vs Regular), and the collaborative behavioral predictors: \ORG{} (\textbf{ORG}), \STR{} (\textbf{STR}), \LE{} (\textbf{LE}), \TP{} (\textbf{TP}), \FP{} (\textbf{FP}). \textbf{Pre Test Average} \rev{and \textbf{Exercise}} were included as control variables to account for each group's initial knowledge \rev{and exercise variability}. All the predictors were evaluated for multi-collinearity using the variance inflation factor (VIF)~\cite{miles2014_VIF} analysis and the values were within acceptable limits.


\subsubsection{Exercise Performance Analysis}  
To compare \gitpushsolved{} across the Split-view and Regular conditions, we first assessed whether the difference scores (Split - Regular) were normally distributed using the Shapiro–Wilk test.
Depending on the outcome, parametric data were analyzed with paired t-tests, and effect sizes were reported as Cohen’s d; non-parametric data were analyzed with Wilcoxon signed-rank tests, with effect sizes calculated as Pearson’s r.

\rev{To ensure that exercise variability did not confound performance comparisons, we also examined whether groups performed differently on \textit{Hangman} versus \textit{Arctic}. As with the condition comparison, we evaluated the normality of within-group difference scores and applied the corresponding paired test with appropriate effect sizes.}

\subsubsection{Comparison of Collaborative Behavior Analysis}  
Differences in the instances of each collaborative behavioral predictor across conditions were analyzed using the same procedure. For each predictor, difference scores were tested for normality with the Shapiro–Wilk test. Parametric data were analyzed with paired t-tests, and effect sizes were reported as Cohen’s d. Non-parametric data were analyzed with Wilcoxon signed-rank tests, with effect sizes calculated as Pearson’s r. P-values from all comparisons were adjusted using the Benjamini–Hochberg procedure to control for inflated Type I error due to multiple testing.

\subsubsection{Post-test Survey Analysis}  
Differences in subjective experiences of cognitive load and perceived collaboration, as measured \rev{respectively} by the two post-test surveys (Sec.~\ref{method:post-test_surveys}), were also assessed using the same procedure. Variables meeting the normality assumption were analyzed with paired t-tests, and effect sizes were reported as Cohen’s d. Variables violating the normality assumption were analyzed with Wilcoxon signed-rank tests, with effect sizes calculated as Pearson’s r.

\subsubsection{Interview Analysis}  
The interview data were transcribed and analyzed using in-vivo coding~\cite{Saldana2013}. One of the authors conducted the initial coding of all interviews, which was subsequently reviewed by a co-author. This process allowed us to familiarize ourselves with participants' perspectives on collaborating to learn Git, their preferences for different aspects of the interfaces and the features they found frustrating.

%% file: sections/06_results.tex
\section{Results}
In this section, we present the key findings of our study, including both the quantitative and qualitative results. 
Overall, our findings show that the \sys{} split-view interface facilitated productive collaborative exchanges for learning Git together in a realistic setting, supported peer learning, fostered social presence and was consistently preferred by participants over the single-view baseline, although overall improvements in Git performance was not observed.

\subsection{Does Git take Two? The Relevance of Collaborative Git Practice}

\subsubsection{Collaborative Practice of Git}

We first investigate the role of collaboration in learning Git by \rev{analyzing how each group’s collaborative behaviors during the exercises shaped their task performance.}
As described in Sec.~\ref{sec:analysis_collab_behavior} a \rev{multiple linear regression} model was fit to predict \gitpushsolved{}, controlling for pre test averages of the groups \rev{and the task variability}. The results \rev{of the regression} are presented in Table~\ref{table:lm}.

\input{tables/06_lm_model}

Our analysis shows that after accounting for \rev{groups'} pre-test averages \rev{and exercise types}, \rev{the final model ($F_{3,22} = 7.987$, $p < .001$, $R^2_{adj} = 0.456$) that significantly predicted \gitpushsolved{} included the pre-test average and two collaborative behaviors predictors.} Irrespective of the assigned study conditions, 
higher \emph{\LE{}} (\textbf{LE}) significantly associated with each groups' greater \gitpushsolved{} ($\beta = 9.\rev{01}$, $SE = 3.\rev{06}$, $t_{22} = 2.\rev{94}$, $p = 0.00\rev{7}$). \rev{The final model also shows association between \emph{\TP{}} (\textbf{TP}) and exercise performance, however it is not significant ($p=0.054$).}
This indicates that the more collaborative learning exchanges \rev{peers engaged in}, the more successful \rev{their} group was in solving the Git tasks.
This result underscores the role of collaboration as a contributing factor to solving Git tasks.

The importance and the need for collaboratively practicing Git is further reflected in participants' interviews. Twenty-four out of the total 26 participants deemed effective coordination and collaborative exchanges as an important aspect of practicing and learning  Git. The key reasons they highlighted for valuing collaborative learning for Git are as follows:
\begin{itemize}[leftmargin=2pt]
    \item[] \textbf{Lowering Barriers.}
    Participants described that collaboratively working with peers help new Git learners bridge the initial fears of “breaking things” and reduces the intimidation of version control systems. Working with peers provides reassurance and builds confidence in making changes. As P01 explained, \inlinequote{For me, it's relevant to collaboratively learn Git because sometimes I feel like pushing something might be a bit scary, and you are scared to change a lot. So if someone else is giving you the `okay', you're less scared to push something.} 
   
    \item[] \textbf{Productivity.}
    Three participants also noted that collaborative Git practice supports productive work by enabling collective ideation, 
    As P03 explained, \inlinequote{I think collaboration definitely helps, when we exchange ideas. But it also helps a lot because when we find the task to be really challenging, it helps to locate problems and find a solution.} Similarly, P04 reflected, \inlinequote{The collaboration part is nice, so we can actually balance ideas.}

    \item[] \textbf{Peer Learning.}
    Furthermore, collaboratively practicing Git gives learners opportunities to observe and exchange Git concepts with their peers.
    P19 expressed, \inlinequote{...it would be helpful to see how others do the task or how others solve the same problems, and may be have the same workflow, and then different people on the team can take steps to make it easier for everyone to work.} Observing how others complete the tasks provides guidance for handling the same tasks or different uses. Also, it provides instant feedback from peers that supports progress when learners encounter difficulties, thereby fostering self-regulated learning. P18 explained, \inlinequote{I do think having another person there allows this sense of feedback loop, and one can maybe tap into the other person's knowledge.}

    \item[] \textbf{Realistic.}
    Especially, participants emphasized that collaborative practice of working on Git mirrors real-world, authentic contexts of distributed version control systems. The collaborative context exposes users to authentic tasks and challenges with peers working on the same repository – like handling merge conflicts and resolving coordination issues. P12 explained, \inlinequote{I think it makes more sense to work with a partner because it's more realistic. The hard part about Git is handling merge conflicts, which only happen when working simultaneously with someone.} Although two participants suggested that these coordination conflicts can also be simulated individually, most participants favoured a collaboratively learning Git since it is more practical and grounded in real-world workflows. P26 explained \inlinequote{...at some point, you will have to work collaboratively with others on Git. And things like pulling while you have changes that are not synced...while there have been additional changes in the remote repository, you wouldn't encounter this when you're working alone...I think it's good to learn this together with someone else.}
\end{itemize}

\subsubsection{Exercise Performance}
Having established the importance of collaborative efforts in practicing Git, we next examined whether the type of collaborative interface influenced participants’ \gitpushsolved{}. A Shapiro–Wilk test indicated that the assumption of normality was not met for the paired differences for \gitpushsolved{}. 
A comparison using the Wilcoxon Signed-rank test of the participants’ \gitpushsolved{} measures between the split-view interface (\md{33.33}, \sd{35.00}) and the single-view interface (\md{66.66}, \sd{36.98}) did not reveal a significant difference ($W = 15.5$, $p = 0.77$, $ns$). 
This does not allow us to validate our H1, however, this outcome is not unexpected. \rev{A paired comparison of performance between the Hangman (\md{33.33}, \sd{37.36}) and Arctic (\md{66.66}, \sd{34.79}) exercises showed no significant difference ($W = 18.50$, $p = ns$, $r = 0.11$), indicating that different exercises did not influence performance. Moreover, because the study operated under a fixed} short completion time, \rev{it could have} likely limited the collaborative learning behaviors from translating into measurable performance gains. 


\rev{Despite no difference in performance, mental effort varied.} When comparing participants' subjective \rev{cognitive load} \rev{using the Wilcoxon signed-rank test}, \rev{the NASA–TLX questionnaire showed \textit{Mental Demand} was} significantly lower ($W=10.50$, $p=0.008$, $r=0.51$) in the split-view condition (\mean{3.53}, \md{4.00}, \sd{1.20}) compared to the regular-view condition (\mean{4.30}, \md{4.00}, \sd{1.37}). \rev{For example, P04 noted that \inlinequote{split-screen view helps in solving these tasks because you or your partner can see quicker where the problem is}, suggesting that the increased awareness reduced the mental effort required to troubleshoot issues collaboratively.}

\subsection{Split-view Interface facilitates Peer Learning for Git}

\input{tables/06_behaviour_comparison}
\begin{figure*}[!htb]
    \centering
    \includegraphics[width=0.9\textwidth]{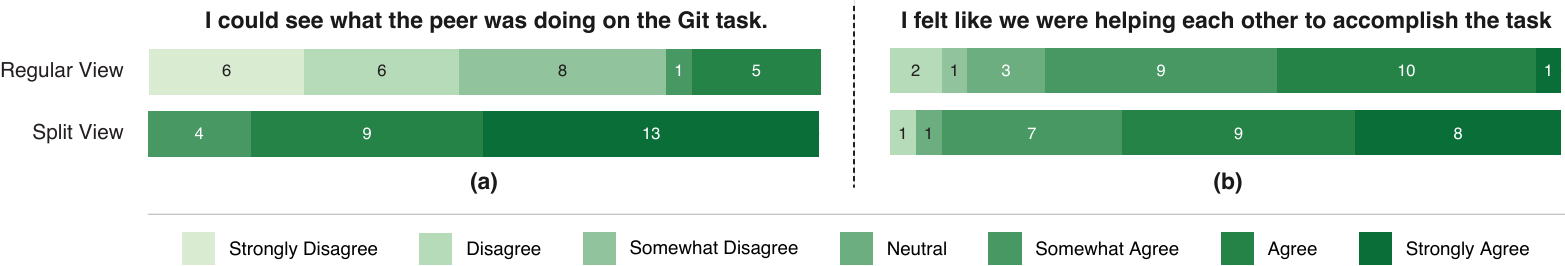}
    \Description{Two horizontal stacked bar charts comparing participant responses in Regular View and Split View conditions. In chart (a), responses to “I could see what the peer was doing on the Git task” show that most Split View responses fall in agreement categories (9 agree, 13 strongly agree), while Regular View responses are concentrated in disagreement categories (6 strongly disagree, 6 disagree, 8 somewhat disagree). In chart (b), responses to “I felt we were helping each other accomplish the task” show stronger agreement in Split View (9 somewhat agree, 8 strongly agree), whereas Regular View responses are more evenly distributed, with many neutral responses (9 neutral).}
    \caption{(a) Participant responses to the social presence survey item: "I could see what the peer was doing on the Git task". (b) Participant responses to the question: "I felt we were helping each other accomplish the task".}
    \label{fig:social_presence}
\end{figure*}

Although \gitpushsolved{} did not differ between interface conditions, we examined whether participants’ collaborative behaviors differed across the conditions \rev{and found that the split-view interface more effectively fostered productive collaboration.}
\

\subsubsection{\rev{Collaborative behaviors in split-view and regular view.}} We compared the collaborative behavior instances across the study conditions and \rev{applying the} Benjamini–Hochberg correction \rev{or multiple comparisons}, we found the following results (Table~\ref{table:behavior_comparison}).
\rev{In split-view, the} instances of \emph{\TP{}} (\textbf{TP}) significantly \rev{differed between conditions} ($t_{12}=2.551$, $p=0.03$, $d=0.57$), with \rev{participants engaging in} more \rev{teaching behaviors} in the split-view condition (\mean{2.61}, \sd{2.32}) compared to regular view (\mean{1.38}, \sd{1.76}).
Similarly, instances of \emph{\FP{}} (\textbf{FP}) was greater in split-view condition (\md{1}, \sd{1.32}) compared to regular view (\md{0.00}, \sd{0.28}), \rev{a} difference was also significant ($W=28$, $p=0.03$, $r=0.65$).


These findings indicate that participants in the split-view interface engaged more frequently in collaborative peer-learning exchanges, either by guiding their collaborators when help was needed or by observing and following their collaborators’ actions to learn and apply similar strategies in their own task solving.

In contrast, participants engaged in significantly more instances of \emph{\ORG{}} (\textbf{ORG}) in the regular view condition (\mean{7.38}, \sd{2.36}) than in the split-view condition (\mean{5.38}, \sd{2.36}), ($t_{12}=-2.79$, $p=0.03$, $d=-0.85$).
Likewise, \emph{\STR{}} (\textbf{STR}) behaviors were higher in the regular view condition (\mean{13.69}, \sd{3.95}) compared to the split-view condition (\mean{8.08}, \sd{4.37}), ($t_{12}=-3.19$, $p=0.03$, $d=-1.35$). This could be attributed to the fact that participants in the regular-view lacked contextual information about their collaborator’s local code states and deictic references. This required them to engage in more back-and-forth communication to organize the task and develop collaborative strategies or resolve conflicts. 
P06 described the difficulty of communicating in the regular-view, \inlinequote{So the regular view condition, it was a bit weird because I didn't know what to communicate. So I was thinking, try to describe the situation because I messed up the reset and then it was hard to really explain what's on my command line.}

\subsubsection{Self-reported collaborative support in split view.}
\rev{Participants’ self-reported experiences also reflected the behavioral differences between the two interface conditions.}
A paired t-test comparing responses to the social presence survey item (Figure~\ref{fig:social_presence}(a)), \inlinequote{I could see what the peer was doing on the Git task}, revealed that participants in the split-view (\mean{6.35}, \sd{0.75}) condition reported greater contextual and visual awareness of their remote peers’ actions compared to those in the regular-view (\mean{2.96}, \sd{1.78}) condition and this difference was significant ($t_{25}=8.62$, $p<0.001$, $d=2.51$). A similar comparison of the responses to the question (Figure~\ref{fig:social_presence}(b)), \inlinequote{I felt we were helping each other accomplish the task}, using a Wilcoxon Signed-rank test showed a significant difference ($W=133.5$, $p<0.01$, $r=0.54$) between the two conditions, with participants reporting higher rating in the split-view (\md{6.00}, \sd{1.17}) compared to the regular-view (\md{5.00}, \sd{1.25}).  

The interviews further indicated that the split-view interface facilitated peer learning by enabling both peer help and observation.
The interface allowed participants to understand their peers’ context accurately and, in turn, share their own context effectively. 
As P23 noted, \inlinequote{I think our levels are a bit different and he requires help. In this case I have to know where he is. And in the split view, he can also see my command, so sometimes I just tell him command that he need to do, and then I just write it to show him.} 
This real-time visibility allowed participants to monitor their peers’ work and \textbf{provide guidance} or \textbf{request help} as needed. 
P8 explained \inlinequote{I was just regularly checking what my peer was doing...I could directly see what she was doing and help quickly or the other way around}. 
Furthermore, \rev{the feedthrough of peers’ actions enabled learners to follow their collaborators’ work}, \textbf{gain insight into their \rev{reasoning and strategies}}, \rev{and use these observations to} guide their own task-solving approaches. P02 quoted, \inlinequote{I like that I could see how she (the collaborator) was moving, that you could see her type on the file, like, I could see also a bit of what her thought process was when working with the task. When she went and checked to see the branches again gave me an idea how she was working with it.} 
Being able to follow peer's process of solving a particular Git concept, also provided users with \rev{\textbf{practical}} \textbf{insight} \rev{\textbf{of how that concept is executed} in a real workflow.} P10 emphasized this point, \inlinequote{I guess watching my partner do the rebasing helped me learn a new concept in Git...it was useful to see how somebody does that.}

\subsection{Features of \sys{}: Enabling Collaborative Git Learning}

Insights from the interviews highlight participants’ preference for the split-screen interface of \sys{}. Among all the participants, 23 indicated that they preferred the split-view interface. Their explanations emphasized specific features of the split-screen view that they found particularly helpful for collaboratively learning and practicing Git. 

\subsubsection{Real Time Mirrored View of Partner's Environment.}
Most participants consistently highlighted the real-time mirrored view of their collaborator’s environment in the split-screen interface of \sys{} as a key advantage. They explained that this feature provided valuable contextual awareness of their partner's workspace. By observing visual feedback on their partner’s code states and actions in real-time, participants gained insight into what their collaborator was experiencing in the task.
P06 described: \inlinequote{So you actually see what he’s typing and going through. That was much more helpful than the single screen where you had to kind of describe what you’re trying to do…I also liked the interaction, the communication, and the visual feedback in the split view.}
    
This mirrored view reduced the need for constant verbal updates and facilitated smoother communication and coordination. P07 similarly noted: \inlinequote{...we kind of had an update without having to explicitly ask for one.}
  
Beyond coordination, participants reported that this awareness allowed them to take initiative and experiment with alternative solutions. As P03 explained: \inlinequote{In the split view, you could directly observe what the other person is doing, so you could, uh, like jump in and attempt stuff or look up some things that could work.}

In addition to supporting effective coordination on Git tasks, the mirrored view of partner's workspace, also fostered a sense of social presence, even though partners where remotely located. P05 emphasized, \inlinequote{When I use the split screen, I do get a feeling that I'm actually working with someone at the same time...even though we are in separate rooms, I feel like he's sitting on my side.} This sense of social presence of working with a partner also made learning of Git less intimidating and more enjoyable, as mentioned by P01, \inlinequote{I guess learning with someone makes it more enjoyable.}
    
\subsubsection{View of Partner's Local Git Tree.}
While the mirrored view of a partner’s working environment afforded general contextual awareness, participants emphasized that the ability to see their partner’s local Git tree made this awareness especially meaningful. Several noted that the Git tree view provided critical insight into branch structures, merges, and repository states, which were otherwise difficult to communicate. For example, P19 highlighted the immediate value of this feature: \inlinequote{Seeing collaborator’s local branches, that’s awesome!} Similarly, P14 explained how the tree view supported task coordination: \inlinequote{The Git tree was more important for me… It helped in some cases, especially when seeing the different merges or different branches on each other’s clients.}
    
P21 further reflected on how the view of local tree enhanced their understanding of peer activity: \inlinequote{In the split screen I felt more like I knew what was going on because I could see the local repository of my partner in real time. In the single screen scenario I sometimes wasn’t sure what he was doing or how his branch looked.} 
    
Furthermore the view of partner's local Git tree also allowed users to verify their own state of work. P25 noted, \inlinequote{The git tree was useful to double check what I was doing and what I had to do.}
Together, these accounts suggest that beyond general surface-level awareness, visibility into the local Git tree gave participants confidence in tracking their collaborator’s workflow, supporting both coordination and mutual learning.
    
\subsubsection{View of Partner's Terminal.}
Alongside the Git tree, participants emphasized the value of being able to see their peer's terminal in real time. The terminal view provided immediate feedback on what commands had been executed, what files were modified, and whether changes had been pushed, enabling collaborators to stay in sync and avoid redundant efforts. As P12 explained, \inlinequote{The terminal was most useful, to know what the other person was entering, what commands, what was already pushed, and what files were modified.}
The ability to scroll through partner's interaction history in the mirrored view of their terminal was also valued by P07, as it allowed them to gain a comprehensive understanding of partner's workflow.
    
For some groups, this awareness even substituted for verbal communication, with collaboration unfolding almost entirely through the shared terminal view. In one case, participants used Vim inside the terminal to chat with each other instead of verbal communication. As P21 noted, \inlinequote{What I think was most useful was to see how my partner worked with the terminal because he's a heavy terminal user…so it was very helpful to see what was going on in his terminal.}
    
Beyond coordination, the terminal view also fostered peer learning by exposing participants to alternative ways of working with Git commands. P16 described how observing their partner's commands broadened their own repertoire: \inlinequote{When I make a new branch, I usually use \texttt{git branch} and then just the name. And he used \texttt{checkout -b}… Things like that are much easier when you see their terminal. And you learn a lot more if you can see what they are doing.}

\subsubsection{View of Partner's File Editor and Mouse Pointer.}
Some participants emphasized that being able to simultaneously view a partner’s file editor and terminal in real time was particularly valuable for collaboration. 
Together, these views enabled peers to coordinate more effectively. As P18 explained, \inlinequote{...the most useful were essentially being able to see the command lines and the like file editing done by the other person. That way, I think it’s a lot easier to conceptualize and to tell like the other person what essentially you think they should click or like how their changes might apply to your team.} 
    
In addition to these core features, a few participants also highlighted the usefulness of deictic cues such as the collaborator’s mouse pointer or cursor, which provided additional context during communication. As P24 described, \inlinequote{The possibility to show things with the mouse pointer in the split view was good} and, P07 explained, \inlinequote{yeah just the context of where the mouse is makes it helpful to understand what she’s saying.}

\subsection{Opportunities for Improvement in \sys{}}
Although users consistently preferred the split view interface of the \sys{}, they also identified opportunities for improvement. 
Several noted that while the split screen supported contextual awareness of their partner's local work environment, the mirrored view reduced the available workspace on their laptop screens. Most participants found the Git tree intuitive, but a few reported confusion with the placement of branch names and icons. Five participants further reported that seeing their partner's mouse pointer added little value to contextual awareness, especially if the partner was a command-line user. Finally, two participants suggested that it would have been useful to see their partner’s activity when searching the Git documentation.

%% file: tables/06_lm_model.tex
{\aptLtoX{\begin{sffont}}{}\begin{table}[ht!]
\centering
\small
\caption{\rev{Final multiple linear regression model predicting \gitpushsolved{} from collaborative behavior predictors, controlling for pre test averages and task. Estimates ($\beta$), standard errors (SE), t-values, and p-values of the predictors in the final model are reported. Significant predictors are indicated by asterisks (*).}}
\label{table:lm}
\rev{
{\aptLtoX{}{\sffamily}
\resizebox{\linewidth}{!}
{%
\begin{tabular}{@{}lllll@{}}
\toprule
\multicolumn{5}{c}{\textbf{Exercise Performance}}                            \\ \midrule
\textbf{Predictor} & \textbf{Estimate ($\beta$)} & \textbf{SE} & \textbf{t-value} & \textbf{p-value} \\
(Intercept)            & -23.72 & 18.03 & -1.315 & 0.2019                    \\
Pre Test Average       & 1.02   & 0.23  & 4.330  & \textbf{\textless 0.001*} \\
Learning Exchange (LE) & 9.01   & 3.06  & 2.943  & \textbf{0.007*}           \\
Teaching Peer (TP)     & -5.59  & 2.74  & -2.038 & 0.054                     \\ \bottomrule
\end{tabular}
}
}%
}
\end{table}\aptLtoX{\end{sffont}}{}}

%% file: tables/06_behaviour_comparison.tex
{\aptLtoX{\begin{sffont}}{}\begin{table*}[ht!]
\centering
\small
\caption{Overview of behavior predictors compared across the split-view and regular view conditions: 
This table reports descriptive statistics – means (M), medians (MD), and standard deviations (SD) for each predictor in the Split-view and Regular conditions. Paired comparisons were conducted using paired t-tests or Wilcoxon signed-rank tests as appropriate. Both raw and Benjamini–Hochberg adjusted p-values are reported, along with effect sizes (Cohen’s d for t-tests and r for Wilcoxon signed-rank tests). Significant findings are highlighted with an asterisk (*).}
\label{table:behavior_comparison}
{\aptLtoX{}{\sffamily}
\begin{tabular}{@{}p{5cm}ccccccp{1cm}p{1cm}p{1.5cm}@{}}
\toprule

 &
 \multicolumn{3}{c}{\textbf{Split View}} & 
 \multicolumn{3}{c}{\textbf{Regular View}} & 
 & & \\
\cmidrule(lr){2-4} \cmidrule(lr){5-7}
\textbf{Predictor} &
  \textbf{\textit{M}} &
  \textbf{\textit{MD}} &
  \textbf{\textit{SD}} &
  \textbf{\textit{M}} &
  \textbf{\textit{MD}} &
  \textbf{\textit{SD}} &
  $\bm{\mathrm{p}_{\mathrm{raw}}}$ &
  $\bm{\mathrm{p}_{\mathrm{adj}}}$ &
  \textbf{\textit{Effect Size}} \\
\addlinespace
\midrule
Organizing (ORG) & 5.38 & 5 & 2.36 & 7.38  & 7  & 2.36 & 0.02 & \textbf{0.03*} & -0.85 \\
Strategizing (STR) & 8.08 & 8 & 4.37 & 13.69 & 13 & 3.95 & 0.01 & \textbf{0.03*} & -1.35 \\
Learning Exchange (LE) & 1.92 & 1 & 2.18 & 1.69 & 2  & 1.55 & 0.65 & 0.65 & 0.12  \\
Teaching Peer (TP) & 2.62 & 2 & 2.33 & 1.38 & 1  & 1.76 & 0.03 & \textbf{0.03*} & 0.57  \\
Following Peer (FP) & 1.31 & 1 & 1.32 & 0.08 & 0  & 0.28 & 0.02 & \textbf{0.03*} & 0.65  \\ 
\bottomrule
\end{tabular}
}
\end{table*}\aptLtoX{\end{sffont}}{}}

%% file: sections/07_discussion.tex
\section{Discussion}


\rev{We found that} participants solved exercises with similar success in both split-view and single-view conditions.
Yet the benefits of split-view surfaced in other important ways: it lowered mental workload, fostered peer teaching and following behaviors, and strengthened social presence.
These findings suggest that immediate performance may not fully capture what matters in Git learning.
Reducing barriers, make distributed state reasoning visible, and supporting peer scaffolding may be more critical outcomes for Git learners than short-term task completion.
In this section, we discuss split view and role play as a pedagogical scaffold for learning Git and its broader implications for collaborative learning. 

\subsection{Why Role Play with Split View is the Right Pedagogical Scaffold for Git}
Git's central challenge is not the syntax of individual commands but the ability to reason about distributed state and coordination across collaborators~\cite{rosso_jackson_2013}.
Most learning resources (e.g., Git Immersion \cite{gitimmersion}, tryGit \cite{tryGit} or GitHowTo \cite{gitHowTo}), however, focus on individual workflows and overlook this coordination layer.
More importantly, they lack the immediacy of working with an actual peer whose actions can disrupt one's own.

Unlike real-time synchronous collaborative editors such as VSCode Liveshare \cite{visualstudioliveShare} or JupyterLab \cite{jupyterlab}, which centralize edits and actions into a single shared view and mask distribution to ensure smooth coordination, Git is inherently distributed with each collaborator maintaining a complete local history that can diverge from others.
This distinction explains why Git learners struggle more with understanding how divergent repositories interact.
In this context, split view is not just a productivity feature but an essential scaffold to surface divergence that real-time editors deliberately hide, and brings a persistent awareness of collaborators' local contexts, crucial to success in distributed collaborative learning~\cite{GutwinEtAl1995, GoswamiEtAl_PD2023}.
Our results suggest that role play with the split-view interface addresses this gap directly. 
By simulating the roles of developers working on Git branches---pulling requests, developing features, merging changes, and resolving conflicts---learners gain hands-on experience with the dynamic responsibilities of team members within collaborative Git workflows.
And, by juxtaposing each learner's local repository with their partner's in real time, GitAcademy externalizes the hidden layer of Git that most novices find confusing: what happens to my state when someone else pushes, merges, or rewrites history? 
\rev{Moreover, learnability of Git is cognitively demanding~\cite{Hynes2015}. Split-view scaffolding helps mitigate this difficulty by allowing learners to follow their peers’ actions, including command sequences, code edits, and resulting state changes as they unfold in real-time. Prior work shows that developers frequently “learn by following” more experienced peers by observing and tracing their work over time~\cite{DabbishEtAl2012}. Similar benefits are documented for novice programmers~\cite{WangEtAl2021}. In this context, split-view offers a structured training scaffold, intentionally providing a comprehensive view of peer's workspace. We recognize this comprehensive visibility to be particularly beneficial for novices, who often need concrete, observable examples to build accurate mental models of Git. Exposing raw code and command output from peers enables learners to see how Git commands produce specific outcomes, internalize effective workflows, and transfer these strategies into their own practice. Even when a peer’s actions do not produce the intended result, the exposure helps learners understand command effects, identify misconceptions, and refine their own workflows.} 

At the same time, our findings also point to new directions for extending this scaffold. 
\rev{In this work, we position this comprehensive awareness through mirrored view of \sys{} as a training scaffold that fades away with progress in learning. Future work could investigate, as learners progress, how the interface might gradually transition to higher-level visual scaffolds---such as summarized workspace differences, simplified coordination cues, or activity awareness dashboards~\cite{BiehlEtAl2007}---that could help interpret distributed states more efficiently.}
Practicing Git collaboratively requires a willing partner, which is not always feasible in classroom or self-study contexts.
Moreover, peers can naturally assume different roles in learning process: sometimes as co-problem-solvers who jointly reason through coordination issues, and at other times as \say{troublemakers} (e.g., pushing to the wrong branch) create authentic challenges for others to solve.
Future work could explore how such roles might be deliberately orchestrated, for example, by integrating AI agents that can take on partner roles, either simulating realistic mistakes or providing just-in-time guidance.
Such hybrid setups could make split-view training more accessible while preserving the benefits of authentic coordination practice.
Even in such AI-augmented scenarios, the split view remains a crucial design element by making the other side of a distributed workflow continuously visible, whether that side belongs to a human collaborator or a simulated one.
Without this perspective, coordination problems are hidden and learners only see the consequences after the fact.


\subsection{Split Views and Role Play in Collaborative Learning of Distributed Workflows}
Although Git was our motivating domain, the design principles we explored extend more broadly of other distributed workflows where reasoning about coordination is as important as executing commands.
We see split-view awareness makes parallel process visible, and role play creates interdependence and asymmetry between learners.

\frev{\textbf{Learning Domains}. Split-view awareness extends to other programming and collaborative learning contexts by supporting visibility into peers' ongoing work without enforcing shared control. 
In classroom-based group projects, such visibility can shift coordination away from turn-taking and explicit handoffs toward parallel planning, timely peer feedback, and early conflict detection, potentially supporting more fluid group dynamics. Through this visibility, learners can adopt and shift between complementary roles—such as planning, checking, explaining, or troubleshooting—based on task demands and momentary expertise rather than predefined assignments. Under time constraints or when expertise is asymmetric among group members, this awareness may reduce coordination overhead and enable more effective knowledge exchange.} 

\frev{This approach is particularly relevant in domains where version control mediates collaborative sense-making rather than code production alone. In data science workflows, for instance, split-view awareness could help team members coordinate when one cleans a dataset while another prepares analysis scripts, making dependencies and potential conflicts visible before they cause pipeline failures. Similarly, in digital humanities and computational research projects, where scholars increasingly use version control to manage collaborative writing, data annotation, or experimental code, simultaneous visibility into collaborators' workspace activities can surface dependencies and evolving project states. In addition to facilitating learner collaboration, split-view functionality could also benefit instructional teams. For example, instructors co-developing teaching materials could use split-view to coordinate changes, delegate tasks, and track the evolution of pedagogical materials.}

\textbf{Technical Domains}. Consider distributed systems work, such as multi-user cloud infrastructures, engineers simultaneously reconfigure machines, networks, and storage.
Each individual sees only their own CLI output or dashboard, but the true difficulty lies in understanding how a teammate's configuration change propagates through the shared system.
A split-view could expose these distributed states in parallel, helping individuals see why coordination protocols and conventions are essential.
When combined with role-play, such as assigning one as the \say{operator} and another as the \say{auditor}, can further scaffold experiential understanding of responsibility, accountability, and recovery in distributed systems.

\textbf{Non-technical Domains}. 
The same principle applies to collaborative work beyond computing such as healthcare teamwork, or crisis response scenarios.
In crisis response scenarios, firefighters, police, and ambulance teams act simultaneously under uncertainty.
A simulator with split view can show how one team's decision (e.g., closing a road) directly affect other's possibilities (e.g., ambulance access).
Structured role assignments would ensure that learners not only see these dependencies but also experience the tension of coordinating.

Together, awareness and role play reinforce one other: learners not only observe a partner's actions but also experience why those actions matter.
This combination creates teachable moments when one peer explains as our study showed.
Cooperative games like It Takes Two \cite{fares2021ittakestwo} demonstrate how deliberately designed interdependence and asymmetry can be engaging rather than frustrating.
We believe that the same design philosophy can translate from entertainment into education, providing learners with authentic, motivating experiences of collaboration in technical domains and beyond.



\subsection{Limitations \& Future Work}

Our study has several limitations. First, the design of \sys{} has areas for improvement, as noted by participants. Future iterations could integrate streaming of partners’ actions on the Git Reference Component to allow for more implicit sharing of their thought processes. Additional enhancements could include tracking and sharing scroll operations, which are not currently implemented. 
Second, the evaluation of our study employed a within-subject lab design with a sample of 26 participants in pairs to assess the effectiveness of \sys{}’s split-screen interface. However, this setup does not allow us to examine the long-term effects of learning at scale through the system. Future studies could replicate this research in real-world educational settings to validate the intervention’s effectiveness. 

%% file: sections/08_conclusion.tex
\section{Conclusion}
Mastering Git extends beyond learning individual commands and concepts; it also requires the ability to reason about distributed states and coordinate actions effectively across collaborators. 
Existing pedagogical approaches often overlook this collaborative dimension. To address this gap, we introduced \sys{}, a browser-based platform for learning Git that scaffolds synchronous collaborative learning and embeds a full Git environment with a split-view collaborative mode: learners work on their own local repositories connected to a shared remote repository, while simultaneously seeing their partner's actions mirrored in real time. 
Our evaluation shows that the split-view interface facilitates peer awareness of distributed processes underlying Git and supports peer learning experiences in Git, even though measurable learning gains were not observed. Our findings have implications for designing training-focused scaffolds to support collaborative learning in Git and other distributed technical systems.

%% file: sections/10_appendix.tex
\clearpage
\onecolumn
\appendix
\section{Appendix}\label{appendix}

\input{tables/A05_demographics_table}
\input{tables/A06_counterbalance_table}
\input{tables/A07_demographic_measures}
\input{tables/A08_git_assessment}
\input{tables/A09_post_test_survey}
\input{tables/A10_interview}

\begin{figure}[!ht]
    \centering
    \includegraphics[width=0.9\columnwidth]{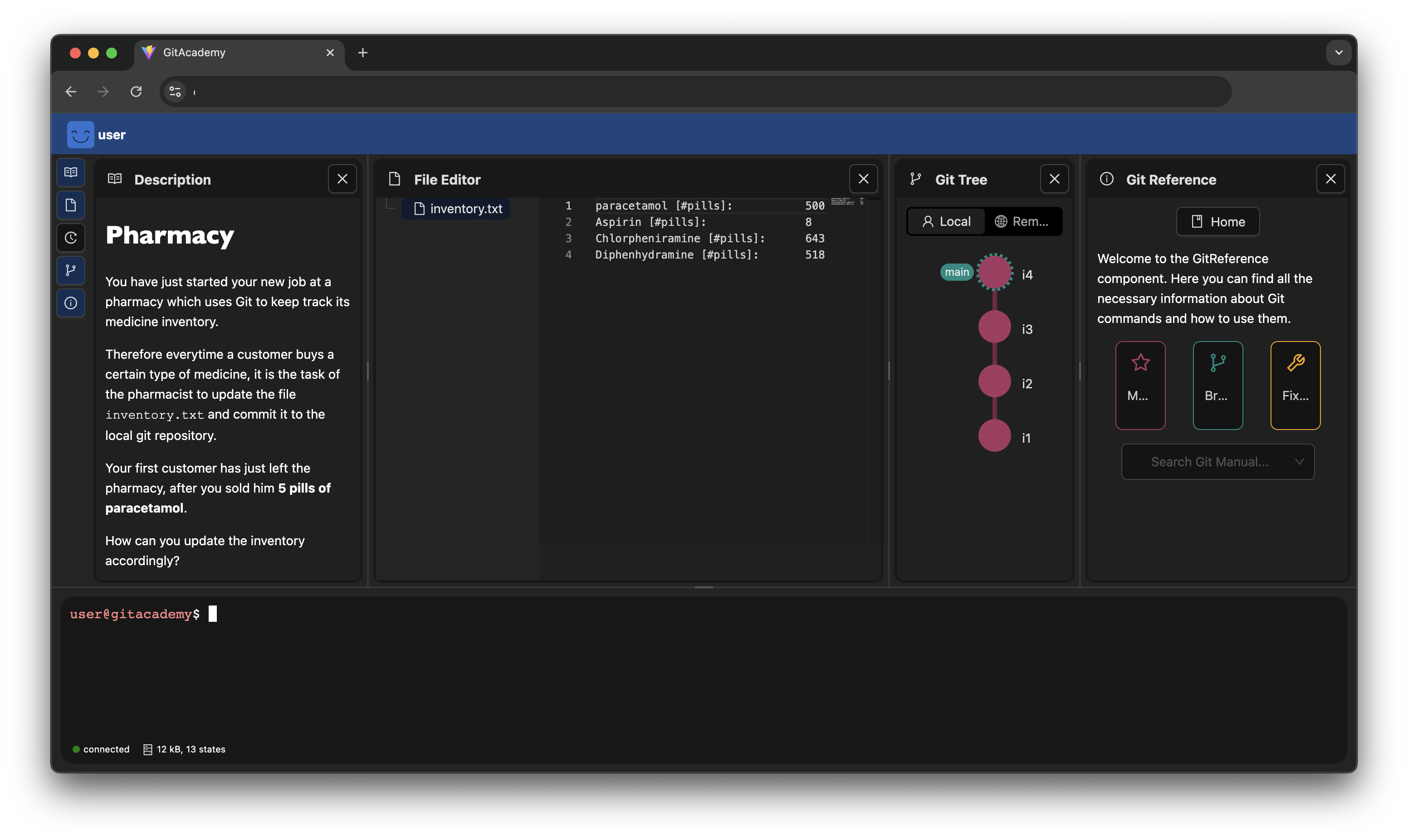}
    \caption{GitAcademy: Screenshot Regular View Exercise View}
    \label{fig:screenshot_regular}
    \Description{Regular view mode of the GitAcademy web-interface showing a single-user exercise. The interface is divided into multiple panels: a task description on the left presenting a pharmacy scenario that asks the learner to update a medicine inventory after selling paracetamol; a file editor in the center displaying an inventory file listing relevant data; a Git tree visualization on the right showing a main branch with several commits; and a Git reference panel providing links to Git commands. A terminal is shown at the bottom with a command prompt, indicating where the learner executes Git commands}
\end{figure}

\begin{figure}[!ht]
    \centering
    \includegraphics[width=0.9\columnwidth]{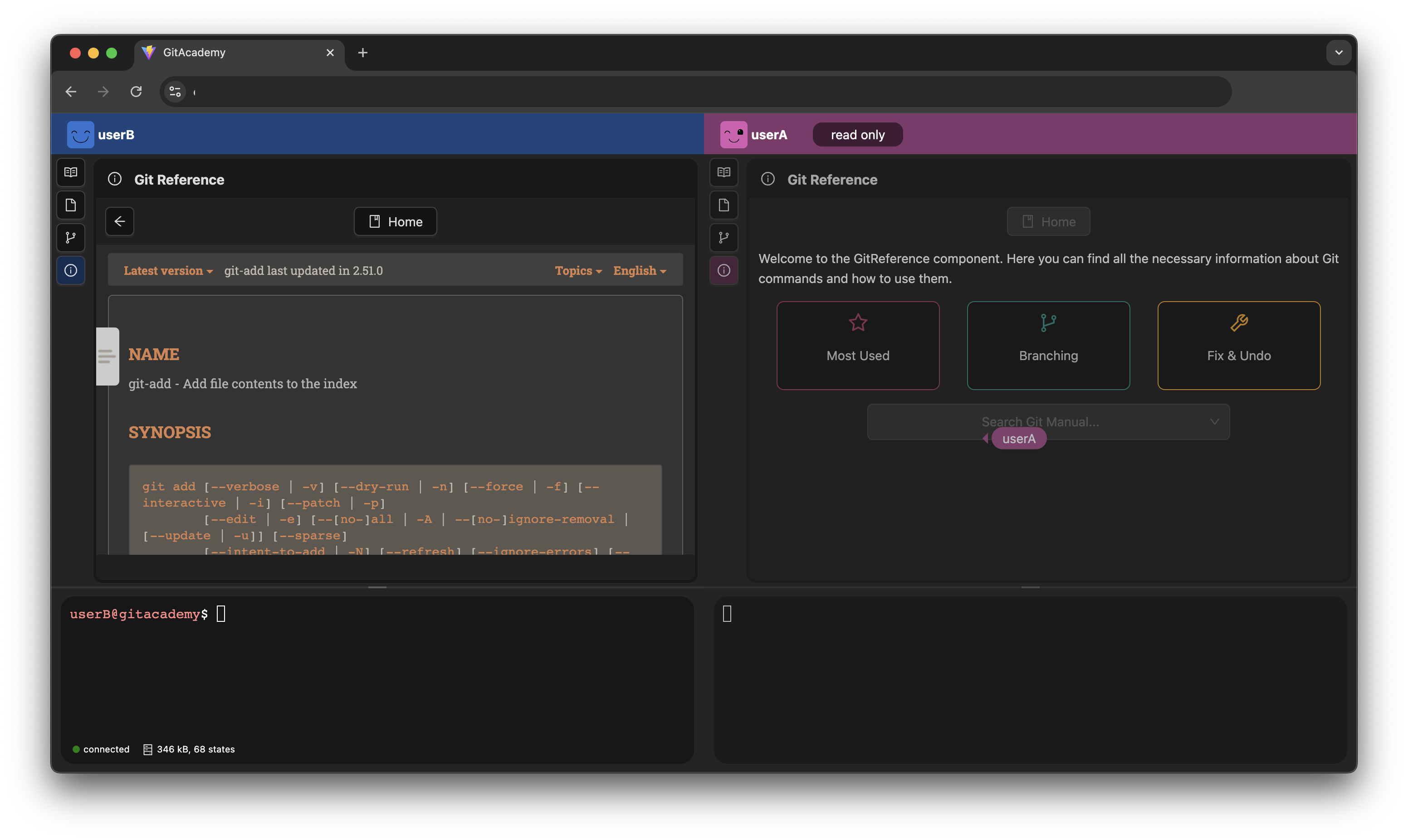}
    \caption{GitAcademy: Screenshot Split View Exercise showing Git Reference Manual}
    \label{fig:screenshot_split_reference}
    \Description{Split-view mode of the GitAcademy interface showing a collaborative Git reference activity. The left panel shows user B’s workspace displaying detailed documentation for the git add command, including its description, usage syntax, and available options. The right panel shows a mirrored, read-only view of user A’s Git Reference workspace, presenting a general overview page with links to common Git topics such as Most Used, Branching, and Fix and Undo. Both panels include a search bar for the Git manual. A terminal window is visible beneath user B’s workspace, indicating where commands are executed, while the mirrored terminal on the right is inactive.}
\end{figure}


%% file: tables/A05_demographics_table.tex
\begin{table}[H]
\centering
\caption{Participants Demographic}
\label{table:demographics}
\begin{tabular}{llllll}
\hline
\textbf{\begin{tabular}[c]{@{}l@{}}Participant \\ ID\end{tabular}} &
  \textbf{Group} &
  \textbf{Age} &
  \textbf{Education} &
  \textbf{\begin{tabular}[c]{@{}l@{}}Git \\ Usage\end{tabular}} &
  \textbf{\begin{tabular}[c]{@{}l@{}}Collaborative \\ Git Usage\end{tabular}} \\ \hline
P01 (F) & 1  & 18 - 24 & Master      & Rarely  & Never   \\
P02 (F) & 1  & 18 - 24 & Master      & Rarely  & Never   \\
P03 (M) & 2  & 25 - 34 & Master      & Weekly  & Rarely  \\
P04 (M) & 2  & 25 - 34 & Master      & Weekly  & Never   \\
P05 (M) & 3  & 25 - 34 & Master      & Monthly & Rarely  \\
P06 (M) & 3  & 25 - 34 & Master      & Rarely  & Rarely  \\
P07 (F) & 4  & 18 - 24 & Master      & Weekly  & Rarely  \\
P08 (F) & 4  & 18 - 24 & Master      & Weekly  & Monthly \\
P09 (M) & 5  & 18 - 24 & Master      & Daily   & Daily   \\
P10 (F) & 5  & 25 - 34 & Master      & Daily   & Rarely  \\
P11 (M) & 6  & 18 - 24 & Master      & Weekly  & Rarely  \\
P12 (F) & 6  & 18 - 24 & Master      & Weekly  & Monthly \\
P13 (M) & 7  & 18 - 24 & Bachelor    & Weekly  & Rarely  \\
P14 (M) & 7  & 25 - 34 & Master      & Weekly  & Rarely  \\
P15 (M) & 8  & 25 - 34 & Master      & Daily   &         \\
P16 (F) & 8  & 25 - 34 & Master      & Weekly  & Monthly \\
P17 (M) & 9  & 18 - 24 & Master      & Daily   & Rarely  \\
P18 (M) & 9  & 18 - 24 & Bachelor    & Weekly  & Rarely  \\
P19 (M) & 10 & 25 - 34 & Master      & Daily   & Rarely  \\
P20 (F) & 10 & 18 - 24 & Master      & Weekly  & Monthly \\
P21 (M) & 11 & 25 - 34 & Master      & Daily   & Monthly \\
P22 (M) & 11 & 18 - 24 & Master      & Daily   & Rarely  \\
P23 (M) & 12 & 25 - 34 & Master      & Daily   & Weekly  \\
P24 (M) & 12 & 18 - 24 & High School & Monthly & Rarely  \\
P25 (M) & 13 & 25 - 34 & Bachelor    & Daily   & Weekly  \\
P26 (X) & 13 & 18 - 24 & Bachelor    & Daily   & Rarely  \\ \hline
\end{tabular}
\end{table}

%% file: tables/A06_counterbalance_table.tex
\begin{table}[H]
\centering
\small
\caption{Counterbalanced conditions for all groups, where Baseline represents regular view, and Experimental represents split view.}
\label{table:counterbalance}
\resizebox{0.6\textwidth}{!}
{%
\begin{tabular}{@{}lllll@{}}
\toprule
\textbf{Group} & \textbf{Task 1} & \textbf{Condition 1} & \textbf{Task 2} & \textbf{Condition 2} \\ 
\midrule
1  & Arctic  & Baseline     & Hangman & Experimental \\
2  & Arctic  & Experimental & Hangman & Baseline     \\
3  & Hangman & Baseline     & Arctic  & Experimental \\
4  & Hangman & Experimental & Arctic  & Baseline     \\
5  & Hangman & Baseline     & Arctic  & Experimental \\
6  & Hangman & Baseline     & Arctic  & Experimental \\
7  & Hangman & Experimental & Arctic  & Baseline     \\
8  & Arctic  & Baseline     & Hangman & Experimental \\
9  & Arctic  & Baseline     & Hangman & Experimental \\
10 & Arctic  & Experimental & Hangman & Baseline     \\
11 & Arctic  & Experimental & Hangman & Baseline     \\
12 & Hangman & Experimental & Arctic  & Baseline     \\
13 & Arctic  & Baseline     & Hangman & Experimental \\ 
\bottomrule
\end{tabular}
}%
\end{table}

%% file: tables/A07_demographic_measures.tex
\begin{table}[H]
\centering
\small
\caption{Demographic Survey Measurements. Answer type: Multiple choice. }
\label{table:demo_measures}
\resizebox{\linewidth}{!} 
{%
\begin{tabular}{ll}
\hline
\textbf{Measurement} &
  \textbf{Choices} \\ \hline
1. What is your age? &
  18-24, 25-34, 35-44, 45-54, 55+ \\
2. What is your gender? &
  Female (F), Male (M), Other (O), Prefer not to say (X) \\
3. What is your highest education level? &
  High School, Bachelor, Master, Ph.D., Other \\
\begin{tabular}[c]{@{}l@{}}4. Which of the following git commands \\ do you feel comfortable to use?\end{tabular} &
  \begin{tabular}[c]{@{}l@{}}git add, git pull, git status, git clone, git push, git commit, \\ git rebase, git init, git branch, git checkout, \\ Resolving merge conflicts, Reverting or resetting commits, \\ Using Git in a team (e.g. handling pull requests)\end{tabular} \\
5. How often do you use Git? &
  Daily, Weekly, Monthly, Rarely, Never \\
\begin{tabular}[c]{@{}l@{}}6. How often do you resolve Git issues \\ that are the result of collaborating with others?\end{tabular} &
  Daily, Weekly, Monthly, Rarely, Never \\ \hline
\end{tabular}
}%
\end{table}

%% file: tables/A08_git_assessment.tex
\begin{table}[H]
\centering
\small
\caption{Git Assessment Measurements. Answer type: Coding challenges.}
\label{table:git_assessment}
\resizebox{\linewidth}{!} 
{%
    \begin{tabular}{l}
        \toprule
         \textbf{Measurement}\\
         \midrule
         1. Can you create a new commit with the commit message \texttt{added new\_file} which contains the file \texttt{new\_file}.\\
         2. You are currently on the branch \texttt{main}. Can you switch to the branch \texttt{dev}. \\
         3. Can you use the terminal to list all branches in your local repository. \\
         4. Can you create a new branch \texttt{branchD}.\\
         5. Can you merge the branch \texttt{dev} into \texttt{main}?\\
         6. Can you rebase the \texttt{dev} branch onto \texttt{main}.\\
         7. Can you tag your current commit with the label \texttt{version1}.\\
         8. Can you merge the branch \texttt{dev} into \texttt{main}. 9. Please ensure that the file \texttt{file.txt} contains (only!) the string \texttt{ABCD} after the merge.\\
         10. Can you stash all your current changes?\\
         \bottomrule
    \end{tabular}
} %
\end{table}

%% file: tables/A09_post_test_survey.tex
\begin{table}[H]
\centering
\caption{Post Exercise Survey Measurements. Answer type: Likert Scale (1 - Very Low or Strongly Disagree, 7 - Very High or Strongly Agree)}
\resizebox{\linewidth}{!} 
{%
    \begin{tabular}{l}
        \toprule
         \textbf{Measurement} \\
         \midrule
         1. How mentally demanding was it to complete the git task when you were using the system? \\
         2. How physically demanding was it to complete the git task when you were using the system?\\
         3. How hurried or rushed was the pace when you were using the system to complete the git tasks?\\
         4. How successful were you in accomplishing the git tasks using the system?\\
         5. How hard did you have to work to accomplish the git task using the system? \\
         6. How insecure, discouraged, irritated, stressed, and annoyed were you when you were using the system to complete the git tasks?\\
         7. I felt like I was working together with the peer to accomplish the git task. \\
         8. I could see what the peer was doing on the git task. \\
         9. I could see that the peer was having difficulties with the git task. \\
         10. I could learn about my peers’ personalities. \\
         11. I could learn about peers’ personal histories.\\
         12. I felt like we were helping each other to accomplish the task \\
         \bottomrule
    \end{tabular}
} %
\label{tab:post_exercise_survey}
\end{table}

%% file: tables/A10_interview.tex
\begin{table}[H]
\centering
\caption{Semi Structured Interview Script}
\resizebox{\linewidth}{!} 
{%
    \begin{tabular}{l}
        \toprule
         \textbf{Interview Questions}\\
         \midrule
         Can you tell me about any past experience you've had with Git before today?\\
         Have you used any tools or platforms specifically designed to teach Git? What worked or didn’t work for you?\\
         Have you ever before collaboratively worked with someone on Git? How was the experience?\\
         How did it feel working with a partner on these Git tasks?\\
         How relevant is it for you to collaborate with a partner for learning/practicing Git concepts? Why?\\
         How is your experience different in two conditions?\\
         Which one of the experiences did you prefer and why?\\
         Which features in the split-screen did you find most useful? Why?\\
         Which features in the split-screen were not useful? Why?\\
         Overall, what part of this experience helped you learn Git the most?\\
         Is there anything you wish the system did differently?\\
         \bottomrule
    \end{tabular}
} %
\label{tab:semi_structured_interview}
\end{table}